\documentclass[article,twocolumn,
amsmath,amssymb,
aps,
prl,
floatfix,
]{revtex4-2}

\usepackage{float}
\usepackage{graphicx}
\usepackage{dcolumn}
\usepackage{bm}
\usepackage{hyperref}
\hypersetup{
    colorlinks=true,
    linkcolor=blue,
    filecolor=magenta,      
    urlcolor=cyan,
}
\usepackage{color}
\usepackage{fancyhdr}
\usepackage[normalem]{ulem}

\usepackage{tikz,xcolor,hyperref}

\definecolor{lime}{HTML}{A6CE39}
\DeclareRobustCommand{\orcidicon}{
	\begin{tikzpicture}
	\draw[lime, fill=lime] (0,0) 
	circle [radius=0.16] 
	node[white] {{\fontfamily{qag}\selectfont \tiny ID}};
	\draw[white, fill=white] (-0.0625,0.095) 
	circle [radius=0.007];
	\end{tikzpicture}
	\hspace{-2mm}
}
\foreach \x in {A, ..., Z}{\expandafter\xdef\csname orcid\x\endcsname{\noexpand\href{https://orcid.org/\csname orcidauthor\x\endcsname}
			{\noexpand\orcidicon}}
}

\begin{document}

\title{Measuring impurity-induced shifts in Coulomb crystallization}

\author{Mingyao Xu}
\affiliation{School of Physics and Astronomy, University of Birmingham, Edgbaston, Birmingham, B15 2TT, United Kingdom}
\author{Aaron A. Smith}
\affiliation{School of Physics and Astronomy, University of Birmingham, Edgbaston, Birmingham, B15 2TT, United Kingdom}
\author{Leonid Prokhorov\orcidB{}}
\affiliation{School of Physics and Astronomy, University of Birmingham, Edgbaston, Birmingham, B15 2TT, United Kingdom}
\author{Vera Guarrera\orcidC{}}
\affiliation{School of Physics and Astronomy, University of Birmingham, Edgbaston, Birmingham, B15 2TT, United Kingdom}
\author{Giovanni Barontini \orcidA{}}
\email{g.barontini@bham.ac.uk}
\affiliation{School of Physics and Astronomy, University of Birmingham, Edgbaston, Birmingham, B15 2TT, United Kingdom}

\date{\today}

\begin{abstract}
We report a laboratory measurement of how impurities shift Coulomb crystallization in a strongly interacting ionic system. This is achieved by using laser cooled Ca$^+$ crystals doped with a controlled number of Xe$^{12+}$ highly charged ions. We find that the crystallization threshold is unchanged at low impurity concentration, but shows a clear crossover once the impurity content becomes sufficiently large, after which the shift grows approximately linearly. Complementary measurements reveal that this global effect originates from a local pinning of the crystal around the impurities. We further show how the measured shift could impact standard models of crystallization in white dwarfs and neutron stars. Our results provide an experimental route to incorporating impurity effects into models of multicomponent Coulomb matter, relevant to stellar crystallization and strongly coupled plasmas.
\end{abstract}

\maketitle

Coulomb crystallization is a first-order transition of strongly coupled ionic matter: a Coulomb liquid freezes once the ratio of inter-ionic potential energy to thermal energy, often parametrized by the Coulomb parameter $\Gamma$, exceeds a critical value (near 175 for the classical one-component plasma) \cite{Hansen1973}. This transition is central to the microphysics of degenerate stellar remnants. In white dwarfs, crystallization and associated phase separation modify the thermal content and energy release of the core, and therefore the cooling history used for cosmochronology \cite{SegretainChabrier1993,Horowitz2010PRL}. In neutron stars, crystallization underpins the existence and mechanical integrity of the solid crust, and strongly influences transport and elasticity that affect observed timing and radiative phenomena \cite{ChamelHaensel2008}.

The Gaia mission has recently revealed prominent features in the local white dwarf cooling sequence consistent with delayed cooling at the onset of core crystallization, and subsequent analyses point to additional composition-driven effects such as buoyant crystals and distillation that can further reshape the predicted cooling evolution \cite{Tremblay2019Nature,Cheng2019ApJ,Blouin2021ApJLNe22,Bedard2024NatureBuoyant}. In these modeling studies, crystallization is typically implemented by adopting an effective critical Coulomb coupling $\Gamma_c$. Impurities are then often treated separately, primarily through their impact on transport via the impurity parameter $Q_{imp} = \langle Z^2\rangle - \langle Z\rangle^2$ (with $Z$ the ionic charge), which quantifies the variance of the ionic charge distribution within the mixture  \cite{Tremblay2019Nature,Cheng2019ApJ,Blouin2021ApJLNe22,Bedard2024NatureBuoyant}. Impurities are central also in neutron star crust formation and evolution: molecular dynamics studies predict strong chemical separation upon freezing, and a pronounced sensitivity of the solidification process to composition \cite{Horowitz2007PRE}. Quantifying how impurities shift $\Gamma_c$ is therefore an open and timely microphysics problem for interpreting Gaia-era crystallization signatures, and for assessing systematic uncertainties in both white dwarf and neutron star modeling. However, impurity effects on Coulomb crystallization remain theoretically challenging because the phase boundary is set by a strongly coupled multicomponent many-body system with long-range interactions, where correlations, disorder, and composition compete in a genuinely non-perturbative regime \cite{Ichimaru1982,DubinONeil1999,Drewsen2015PhysicaB,Kozhberov2024,ChamelHaensel2008}.

Despite the enormous disparity in density, temperature, and length scale, cold trapped ion Coulomb crystals and dense stellar Coulomb matter are governed by the same underlying physics. They are both systems of charged particles with long-range Coulomb interactions, whose structural properties are controlled primarily by the competition between interaction energy and thermal motion, parametrized by $\Gamma$. Cold trapped ion crystals therefore provide a realization of the same strongly coupled Coulomb plasma regime relevant to white dwarf interiors and neutron star crusts, but in controlled tabletop settings. Laser cooled ions in radiofrequency traps realize Coulomb crystals with tunable $\Gamma$ and well defined composition, and their spatial order can be quantified directly by fluorescence imaging. Mixed-species crystals are also routinely produced via sympathetic cooling \cite{PhysRevLettxx.105.143001,chou2017preparation,krohn2023reactions,Drewsen2015PhysicaB,Blackburn2020SciRep,Blackburn2025NJP,khanyile2015observation,rugango2015sympathetic,okPhysRevApplied.4.054009,petralia2020strong,guggemos2015sympathetic}. Recent advances have extended this capability to highly charged ions \cite{Schmoger2015Science,ChenShaolong2025,prokhorov2025coulombcrystallizationxenonhighly}, enabling controlled high-$Z$ dopants that closely mimic the role of impurities in dense stellar Coulomb plasmas. The combination of tunable coupling, controlled doping, and direct structural diagnostics therefore makes trapped ion crystals a powerful platform for quantitatively measuring how impurities modify Coulomb crystallization, and for providing microphysics inputs that could be incorporated into plasma and stellar models.

\begin{figure*}
\centering
\includegraphics[width=0.9\textwidth]{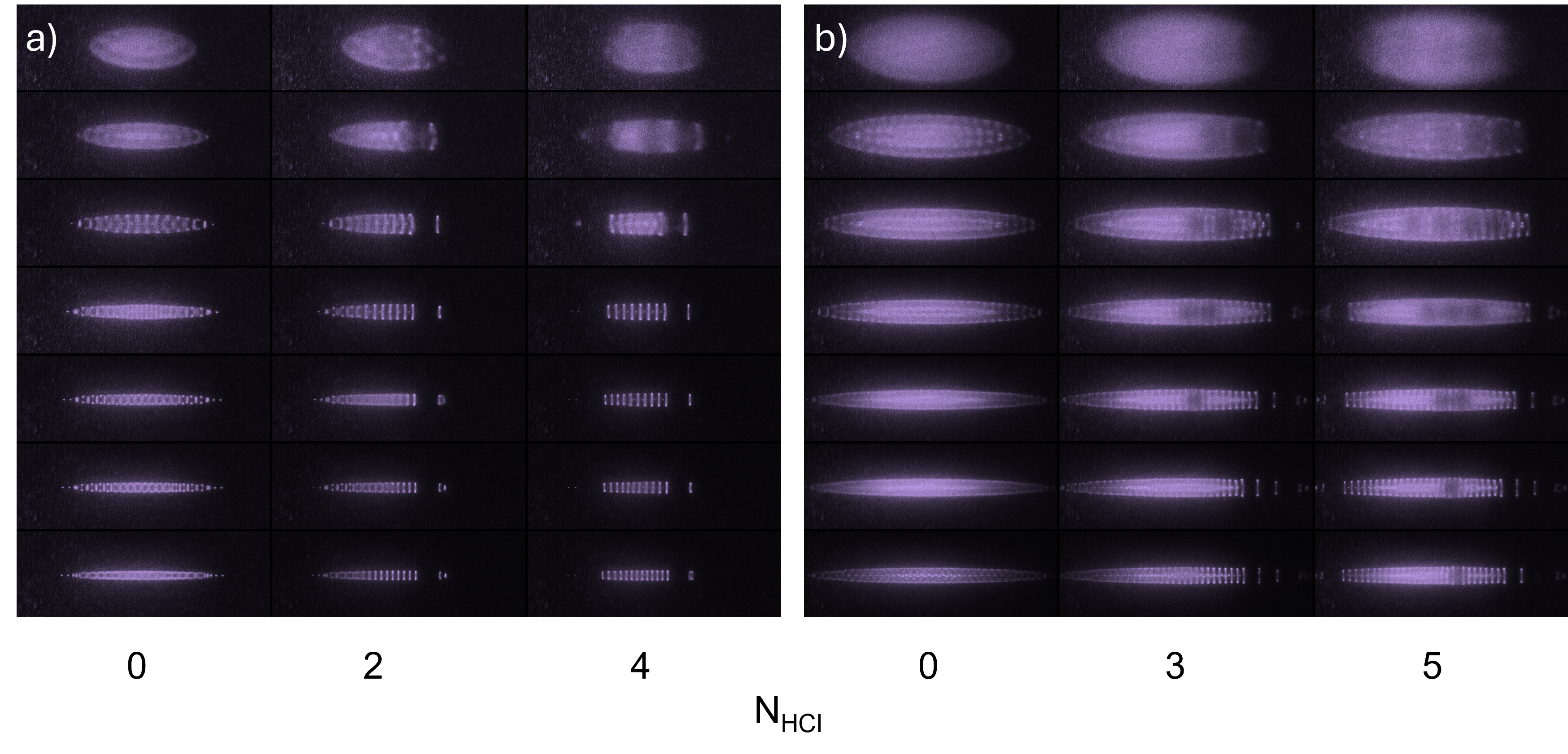}
\caption{a) Fluorescence images of samples of 50 $\pm$ 5 Ca$^+$ ions across the liquid-to-crystal crossover. The observable voids are caused by Xe$^{12+}$ ions, acting as dopants. The number of Xe HCIs increases from left to right (the first column is the undoped reference). The trapping frequencies in the first row are $(\omega_x,\omega_y,\omega_z)=2\pi\times(192,262,164)$ kHz. By increasing the radial confinement (from top to bottom), corresponding to increasing the Coulomb parameter $\Gamma$, we observe the progressive onset of crystallization. The radial frequencies in the last row are $(\omega_x,\omega_y)=2\pi\times(888,897)$ kHz. Each image is $\simeq330\times110$ $\mu$m, and the exposure time is 300 ms. b) The same as a) but with 191 $\pm$ 23 Ca$^+$ ions.}
\label{Fig1}
\end{figure*}

In this Letter we report a laboratory measurement of impurity-induced shifts in Coulomb crystallization. Using laser-cooled Ca$^+$ Coulomb crystals doped with a controlled number of Xe$^{12+}$ highly charged ions, we determine how $\Gamma_c$ depends on the impurity content. We find that the crystallization threshold is essentially unchanged at small $Q_{imp}$, but shows a clear crossover once $Q_{imp}\simeq O(1)$; beyond this point, $\Gamma_c$ decreases approximately linearly with $Q_{imp}$. By combining measurements in large ensembles with complementary studies in small, ion-resolved crystals, we connect this global renormalization of the crystallization boundary to local structural changes. Finally, we illustrate how the measured $\Gamma_c(Q_{imp})$ can be propagated into representative white dwarf and neutron star crystallization prescriptions, yielding substantial shifts in predicted crystallization onset and other structural signatures. Our work therefore provides useful indications for plasma physics and stellar applications.

Details of our experimental system are provided in \cite{prokhorov2025coulombcrystallizationxenonhighly}. In brief, Xe$^{q+}$ bunches are produced in a compact EBIT, charge-selected ($q=12$ for this work), decelerated, and transported with electrostatic optics into a cryogenic (4 K) science chamber hosting a segmented linear Paul trap. In the same trap we prepare Ca$^+$ Coulomb crystals using Doppler cooling. The Ca$^+$ fluorescence is then imaged onto an EMCCD camera with a 30$\times$ magnification. The incoming Xe$^{12+}$ ions are sympathetically cooled through repeated interactions with the preformed Ca$^+$ crystal until they crystallize. Each embedded HCI appears as a localized dark void in the Ca$^+$ fluorescence. In this work, we deterministically prepare mixed-species Coulomb crystals containing a prescribed number of Ca$^+$ ions and a chosen number (1–6) of embedded Xe$^{12+}$ ions. Efficient capture and sympathetic cooling of the HCIs requires a large preformed Ca$^+$ crystal. We therefore first load a crystal containing hundreds of Ca$^+$ ions, and then implant Xe$^{12+}$ repeatedly until the desired number is reached. To then reduce the Ca$^+$ ion number to the target value, we first fully melt the crystal by decreasing the radial confinement, and then blue-detune and attenuate the cooling light. In this way the Ca$^+$ ions are progressively expelled. By monitoring the decrease of the total fluorescence, we obtain single-ion resolution on the remaining Ca$^+$ number. After this preparation, the laser parameters are set back to the optimal operating point.

To study the liquid-to-crystal transition, after preparation we increase the radial confinement of the Paul trap. This increases the ion density and hence the Coulomb coupling parameter $\Gamma = e^2/(4\pi\varepsilon_0 a k_B T)$, where $a$ is the Wigner–Seitz radius set by the local Ca$^+$ density, $T$ the temperature, $e$ the elementary charge, $\varepsilon_0$ the vacuum permittivity, and $k_B$ the Boltzmann constant. In this work we use the single-component Coulomb parameter $\Gamma$ as the central metric, and quantify doping effects through the resulting shift of the Ca$^+$ liquid-to-solid transition \cite{SM}. As the confinement is ramped up, the ensemble evolves from a strongly correlated Coulomb liquid to an ordered crystal. We study Ca$^+$ samples containing between 20 and 350 ions. For a fixed number of Ca$^+$ ions, we ramp across the transition varying the number of implanted Xe$^{12+}$ ions. For each set of parameters, we collect 5 fluorescence pictures (exposure time of 300 ms). In Fig. 1 we show an example of our protocol for a Ca$^+$ sample of 50 ions with up to 4 Xe$^{12+}$ ions implanted, and one of 191 ions with up to 5 HCIs implanted. 

Traditionally, the onset of crystallinity in strongly coupled ionic matter is characterized by the Lindemann parameter $\mathcal{L} = \sqrt{\langle u^2\rangle}/a$, where $\langle u^2\rangle$ is the mean-squared displacement about equilibrium sites. Melting is typically associated with $\mathcal{L}$ of order 0.1. For large three-dimensional crystals, however, $\mathcal{L}$ is not directly accessible from fluorescence images because individual ions are not resolved, and most of them lie outside the focal plane. Extracting $\mathcal{L}$ then becomes indirect and model-dependent, requiring assumptions about the density profile, mapping from image blur to spatial fluctuations, and related systematics. Here we use an image-based metric, the Tenengrad (squared-gradient) sharpness, to quantify the emergence of crystalline order. Briefly, for each picture we compute the Sobel derivatives and form the Tenengrad energy map $\mathcal{T}(x,y)$. We then compute the sharpness of the picture by averaging $\mathcal{T}$ over the sample, yielding a single Tenengrad value per frame, see \cite{SM} for details. We show in \cite{SM} that, within a simple imaging model, $\langle\mathcal{T}\rangle$ is a decreasing function of $\mathcal{L}$, and therefore provides a quantitative measure of crystallinity.

\begin{figure}
\centering
\includegraphics[width=0.48\textwidth]{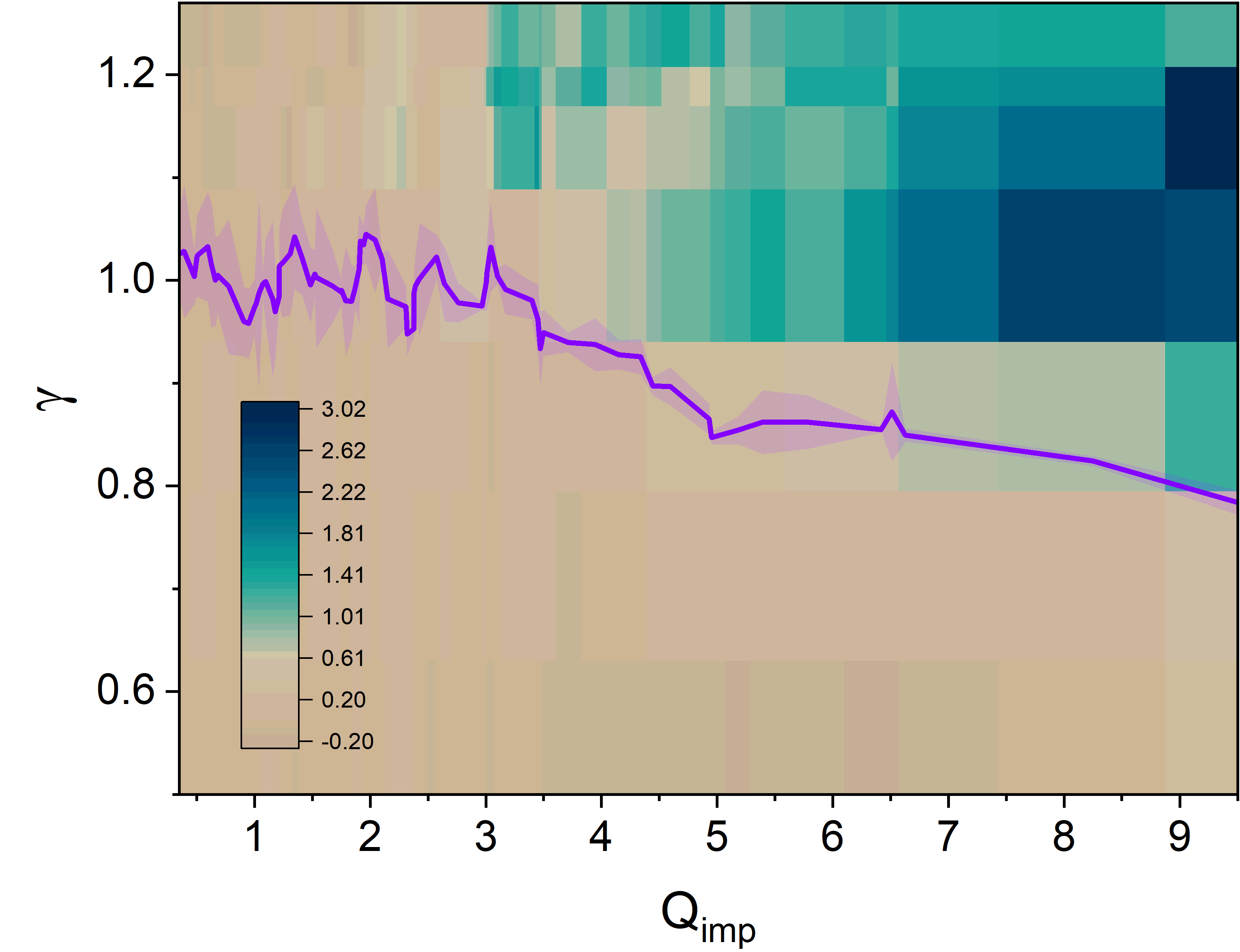}
\caption{The purple line is the measured value of the normalized critical Coulomb parameter $\gamma=\Gamma_c(Q_{imp})/\Gamma_c(0)$ as a function of the impurity parameter $Q_{imp}$. The shaded area corresponds to the standard deviation \cite{SM}. The color scale of the underlying contour plot is the normalized Tenengrad deviation $\delta=\Delta\langle \mathcal{T}\rangle/\langle \mathcal{T}(0)\rangle$.}
\label{Fig2}
\end{figure}

Because our system is highly inhomogeneous, it does not feature a sharp phase transition but a smooth crossover between the liquid and the crystal phases. We therefore define a reference crystallinity threshold from the undoped Ca$^+$ data as the midpoint between the low-confinement response (in the example of Fig.~1, the first row) and the typical high-confinement response (last rows in Fig.~1). For each dopant concentration, we construct the discrete curve $\langle\mathcal{T}\rangle(\Gamma)$ and extract $\Gamma_c$ as the value of $\Gamma$ at which $\langle\mathcal{T}\rangle(\Gamma)$ crosses this threshold, using linear interpolation between neighboring points. Additional details are provided in \cite{SM}. In Fig. 2 we report the measured dependence of $\gamma=\Gamma_c(Q_{imp})/\Gamma_c(0)$ on the impurity parameter $Q_{imp}$. We observe that the transition is not affected by the dopant until $Q_{imp}$ becomes of order unity. At $Q_{imp}\simeq 2.1$, $\gamma$ exhibits a knee, and starts to decrease linearly with $Q_{imp}$ \cite{SM}. This behavior is confirmed by the normalized Tenengrad deviation, $\delta=\Delta\langle \mathcal{T}\rangle/\langle \mathcal{T}(0)\rangle$, where $\Delta\langle \mathcal{T}\rangle \equiv \langle \mathcal{T}(Q_{imp})\rangle-\langle \mathcal{T}(0)\rangle$. By construction, $\delta$ directly quantifies how much the inferred level of crystallinity differs from the pure Ca$^+$ sample in the same conditions. The color scale in Fig. 2 shows that $\delta$ starts to significantly deviate from zero for large values of $\gamma$ when $Q_{imp}$ becomes larger than one. It then increases systematically as $Q_{imp}$ is raised. At the same time, $\delta$ deviates from zero for increasingly smaller values of $\gamma$ as $Q_{imp}$ increases, confirming the behavior previously discussed.

By analyzing the spatial distribution of $\mathcal{T}(x,y)$, we observe that, along the main axis of the sample, the spatial range over which an embedded HCI perturbs the Ca$^+$ crystal is $50\pm5$ $\mu$m below $\Gamma_c$, and increases to $60\pm5$ $\mu$m above $\Gamma_c$ \cite{SM}. To gain further insight into the local effect of the impurity, we prepare smaller crystals (up to 20 Ca$^+$ ions) in which individual ions are reliably resolved, allowing an estimate of the Lindemann parameter \cite{SM}. In this regime the global inhomogeneity is reduced and local structural changes around the dopant can be quantified without relying on image-based methods. Also in this case, we control the crossover to crystallization by varying the radial confinement. In these small samples this plays the role of an effective pressure set by the surrounding ions in large crystals. For convenience, we parameterize the confinement by the normalized radial “pressure” $\Pi = (\omega_r/\omega_r^{\mathrm{l}})^2$, where $\omega_r^{\mathrm{l}}$ is the radial secular frequency at which the undoped sample is in the liquid phase. Our results are reported in Fig. 3 and show that $\mathcal{L}$ depends strongly on the impurity content. The $\mathcal{L} = 0.1$ line, which approximately sets the transition from liquid to crystal, shifts to substantially lower $\Pi$ as $Q_{imp}$ increases. In other words, for fixed radial pressure, the ions are more localized in the presence of dopants, indicating a pronounced local pinning effect of the HCIs that modifies the local melting criterion, similar to what was observed in \cite{Rueffert2024}. Taken together, our data indicate that a dopant primarily acts locally by stabilizing an ordered region around itself. In large crystals, the global shift of $\Gamma_c$ emerges once these locally pinned regions extend over a non-negligible fraction of the host crystal, consistent with the onset for $Q_{imp}> 1$.

\begin{figure}
\centering
\includegraphics[width=0.48\textwidth]{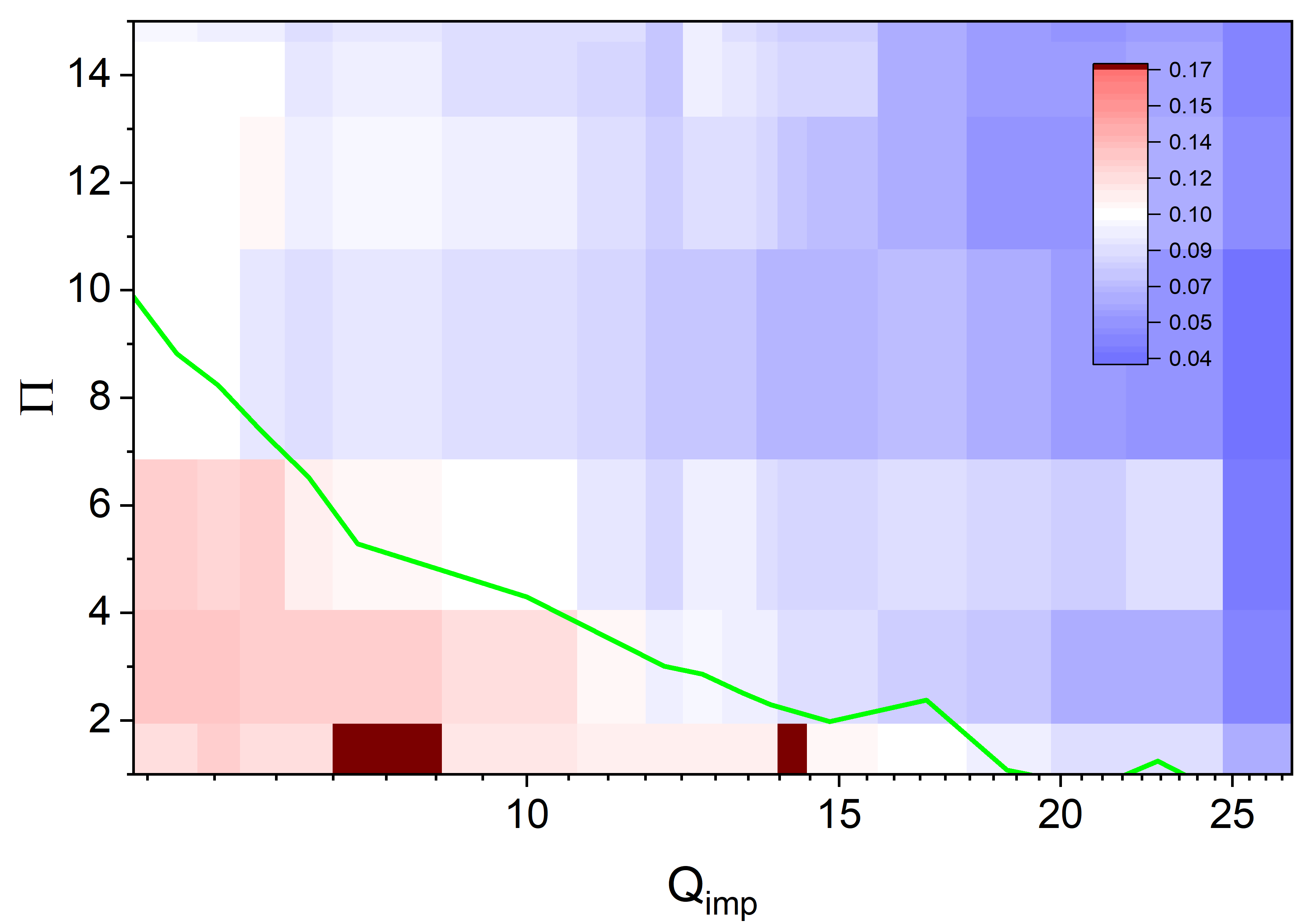}
\caption{Lindemann parameter $\mathcal{L}$ as a function of the normalized radial pressure $\Pi$ and the impurity parameter $Q_{imp}$. The green line corresponds to the condition $\mathcal{L}=0.1$, which approximately sets the boundary between the liquid and the crystal phases.}
\label{Fig3}
\end{figure}

To provide a crude estimation of how our measured shift might impact the crystallization in white dwarfs, we fit $\gamma(Q_{imp})$ with a phenomenological knee function \cite{SM}, and then we incorporate it as a correction to the freezing condition of a strongly coupled ionic plasma in a degenerate electron background. At fixed density, $\Gamma \propto 1/T$ implies a local melting temperature shift
$T_m(Q_{imp})/{T_m(0)}=1/{\gamma(Q_{imp})}$. Using the standard Mestel scalings for a degenerate core with an insulating envelope \cite{Mestel1952,ShapiroTeukolsky,Bhattacharya2018MNRAS_Mestel}, we obtain that the luminosity scales as ${L(Q_{imp})}/{L(0)}=1/\gamma(Q_{imp})^{7/2}$, while the corresponding shift in the crystallization time scales as
$t_{\rm crys}(Q_{imp})/{t_{\rm crys}(0)}=\gamma(Q_{imp})^{5/2}$ \cite{SM}. Therefore, impurity-induced changes in $\gamma$ are amplified into large shifts in the inferred luminosity and crystallization age, as shown in Fig. 4a and 4b. In neutron-star crust modeling instead, at fixed temperature the crystallization density shifts as $\rho_m(Q_{imp})/\rho_m(0)=\gamma(Q_{imp})^3$ \cite{ChamelHaensel2008,HaenselPotekhinYakovlev2007}, also leading to a substantial change, as shown in Fig. 4c. This corresponds to a displacement of the solid–liquid boundary, and therefore to a change of the depth range over which elastic (shear) and solid state transport properties apply. Note that these scalings apply in the Coulomb limit of weak screening; for finite screening one would need to replace the one-component plasma criterion with the Yukawa melting curve and then apply the same corrections \cite{Hamaguchi1997}.

In conclusion, we have used Ca$^+$ Coulomb crystals doped with Xe$^{12+}$ ions to measure an impurity-induced renormalization of the liquid-to-solid transition. We have developed a simple method to quantitatively analyze fluorescence images of large Coulomb crystals. Our data show that the normalized threshold $\gamma(Q_{imp})$ is essentially unchanged for low dopant concentrations, and exhibits a clear crossover when the impurity parameter becomes of order unity. It then decreases approximately linearly with the impurity content. By utilizing ion-resolved measurements, we showed that the global shift is rooted in a local pinning mechanism, with an influence range of tens of micrometers. We have shown that the shift measured in this work could result in large corrections to several stellar observables, due to the rapid scaling. This, for example, could provide some indications for interpreting crystallization signatures recently observed by the Gaia mission, where the cooling sequence shows features consistent with delayed cooling at the onset of core crystallization, and additional composition effects \cite{Tremblay2019Nature,Cheng2019ApJ,Blouin2021ApJLNe22,Bedard2024NatureBuoyant}. A detailed implementation in stellar evolution calculations will require going beyond the simplifying assumptions used here (e.g. including screening, multicomponent phase behavior, and transport), but our measurement could provide a concrete microphysics input for the impurity sensitivity of the phase boundary itself. More broadly, impurity-controlled crystallization is a generic problem in strongly coupled plasma physics. Our setup and method provide a controlled benchmark for impurity physics in strongly coupled Yukawa/Coulomb systems, and can be used to test and calibrate models of dopant-induced defect formation, pinning, and disorder, that are central in complex plasma studies. Looking ahead, an interesting route is to quantify impurity transport by applying controlled electric fields and measure drift and diffusion across the crossover. Another promising avenue is to investigate impurity-impurity interactions mediated by the collective modes of the host crystal, connecting naturally to polaron-type physics \cite{Stojanovic2012PRL}.

\begin{figure}
\centering
\includegraphics[width=0.48\textwidth]{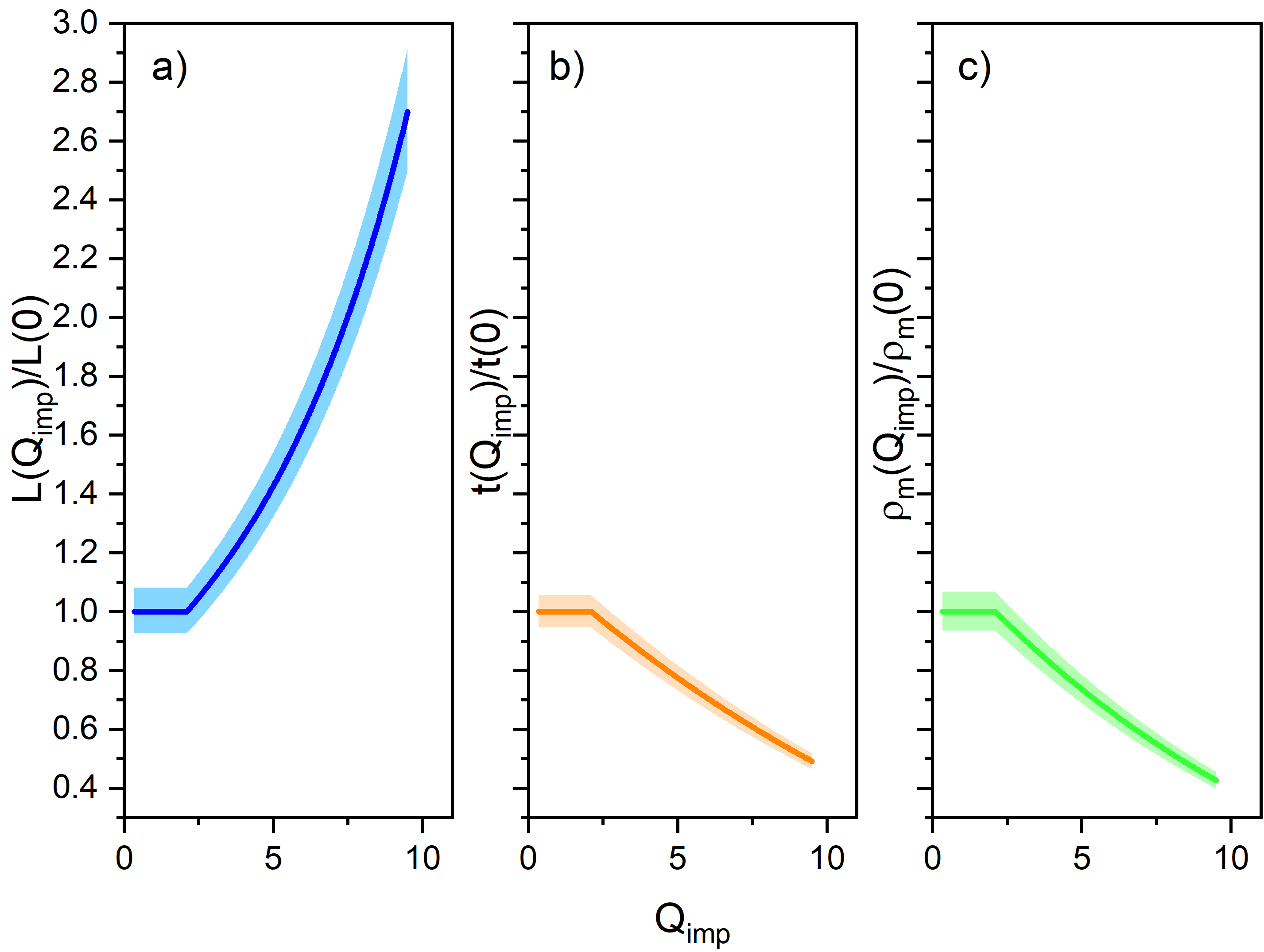}
\caption{Evaluation of the potential impact of the impurity-induced shift of crystallization measured in this work on stellar models. a) Relative shift in the luminosity at crystallization onset in white dwarfs. b) Corresponding shift in the crystallization age. (c) Shift in the crystallization density in the neutron-star crust at fixed temperature. The shaded areas correspond to the average uncertainty in $\gamma$ \cite{SM}.}
\label{Fig4}
\end{figure}

\paragraph*{Acknowledgements}
We are grateful to Xen Sergi for technical support. We acknowledge fruitful discussions with the members of the Atomic Quantum Systems group at the University of Birmingham. This work was supported by STFC and EPSRC under grants ST/Y00454X/1, ST/W006138/1, and ST/T00603X/1.  

\bibliography{main_bibl}

\begin{thebibliography}{40}%
\makeatletter
\providecommand \@ifxundefined [1]{%
 \@ifx{#1\undefined}
}%
\providecommand \@ifnum [1]{%
 \ifnum #1\expandafter \@firstoftwo
 \else \expandafter \@secondoftwo
 \fi
}%
\providecommand \@ifx [1]{%
 \ifx #1\expandafter \@firstoftwo
 \else \expandafter \@secondoftwo
 \fi
}%
\providecommand \natexlab [1]{#1}%
\providecommand \enquote  [1]{``#1''}%
\providecommand \bibnamefont  [1]{#1}%
\providecommand \bibfnamefont [1]{#1}%
\providecommand \citenamefont [1]{#1}%
\providecommand \href@noop [0]{\@secondoftwo}%
\providecommand \href [0]{\begingroup \@sanitize@url \@href}%
\providecommand \@href[1]{\@@startlink{#1}\@@href}%
\providecommand \@@href[1]{\endgroup#1\@@endlink}%
\providecommand \@sanitize@url [0]{\catcode `\\12\catcode `\$12\catcode `\&12\catcode `\#12\catcode `\^12\catcode `\_12\catcode `\%12\relax}%
\providecommand \@@startlink[1]{}%
\providecommand \@@endlink[0]{}%
\providecommand \url  [0]{\begingroup\@sanitize@url \@url }%
\providecommand \@url [1]{\endgroup\@href {#1}{\urlprefix }}%
\providecommand \urlprefix  [0]{URL }%
\providecommand \Eprint [0]{\href }%
\providecommand \doibase [0]{https://doi.org/}%
\providecommand \selectlanguage [0]{\@gobble}%
\providecommand \bibinfo  [0]{\@secondoftwo}%
\providecommand \bibfield  [0]{\@secondoftwo}%
\providecommand \translation [1]{[#1]}%
\providecommand \BibitemOpen [0]{}%
\providecommand \bibitemStop [0]{}%
\providecommand \bibitemNoStop [0]{.\EOS\space}%
\providecommand \EOS [0]{\spacefactor3000\relax}%
\providecommand \BibitemShut  [1]{\csname bibitem#1\endcsname}%
\let\auto@bib@innerbib\@empty
\bibitem [{\citenamefont {Hansen}(1973)}]{Hansen1973}%
  \BibitemOpen
  \bibfield  {author} {\bibinfo {author} {\bibfnamefont {J.~P.}\ \bibnamefont {Hansen}},\ }\bibfield  {title} {\bibinfo {title} {Statistical mechanics of dense ionized matter. i. equilibrium properties and melting transition of the one-component plasma},\ }\href {https://doi.org/10.1103/PhysRevA.8.3096} {\bibfield  {journal} {\bibinfo  {journal} {Physical Review A}\ }\textbf {\bibinfo {volume} {8}},\ \bibinfo {pages} {3096} (\bibinfo {year} {1973})}\BibitemShut {NoStop}%
\bibitem [{\citenamefont {Segretain}\ and\ \citenamefont {Chabrier}(1993)}]{SegretainChabrier1993}%
  \BibitemOpen
  \bibfield  {author} {\bibinfo {author} {\bibfnamefont {L.}~\bibnamefont {Segretain}}\ and\ \bibinfo {author} {\bibfnamefont {G.}~\bibnamefont {Chabrier}},\ }\bibfield  {title} {\bibinfo {title} {Crystallization of binary ionic mixtures in dense stellar matter},\ }\href@noop {} {\bibfield  {journal} {\bibinfo  {journal} {Astronomy and Astrophysics}\ }\textbf {\bibinfo {volume} {271}},\ \bibinfo {pages} {L13} (\bibinfo {year} {1993})}\BibitemShut {NoStop}%
\bibitem [{\citenamefont {Horowitz}\ \emph {et~al.}(2010)\citenamefont {Horowitz}, \citenamefont {Schneider},\ and\ \citenamefont {Berry}}]{Horowitz2010PRL}%
  \BibitemOpen
  \bibfield  {author} {\bibinfo {author} {\bibfnamefont {C.~J.}\ \bibnamefont {Horowitz}}, \bibinfo {author} {\bibfnamefont {A.~S.}\ \bibnamefont {Schneider}},\ and\ \bibinfo {author} {\bibfnamefont {D.~K.}\ \bibnamefont {Berry}},\ }\bibfield  {title} {\bibinfo {title} {Crystallization of carbon-oxygen mixtures in white dwarf stars},\ }\href {https://doi.org/10.1103/PhysRevLett.104.231101} {\bibfield  {journal} {\bibinfo  {journal} {Physical Review Letters}\ }\textbf {\bibinfo {volume} {104}},\ \bibinfo {pages} {231101} (\bibinfo {year} {2010})}\BibitemShut {NoStop}%
\bibitem [{\citenamefont {Chamel}\ and\ \citenamefont {Haensel}(2008)}]{ChamelHaensel2008}%
  \BibitemOpen
  \bibfield  {author} {\bibinfo {author} {\bibfnamefont {N.}~\bibnamefont {Chamel}}\ and\ \bibinfo {author} {\bibfnamefont {P.}~\bibnamefont {Haensel}},\ }\bibfield  {title} {\bibinfo {title} {Physics of neutron star crusts},\ }\href {https://doi.org/10.12942/lrr-2008-10} {\bibfield  {journal} {\bibinfo  {journal} {Living Reviews in Relativity}\ }\textbf {\bibinfo {volume} {11}},\ \bibinfo {pages} {10} (\bibinfo {year} {2008})}\BibitemShut {NoStop}%
\bibitem [{\citenamefont {Tremblay}\ \emph {et~al.}(2019)\citenamefont {Tremblay}, \citenamefont {Fontaine}, \citenamefont {Gentile~Fusillo}, \citenamefont {Dunlap}, \citenamefont {G{\"a}nsicke}, \citenamefont {Hollands}, \citenamefont {Hermes}, \citenamefont {Marsh}, \citenamefont {Cukanovaite},\ and\ \citenamefont {Cunningham}}]{Tremblay2019Nature}%
  \BibitemOpen
  \bibfield  {author} {\bibinfo {author} {\bibfnamefont {P.-E.}\ \bibnamefont {Tremblay}}, \bibinfo {author} {\bibfnamefont {G.}~\bibnamefont {Fontaine}}, \bibinfo {author} {\bibfnamefont {N.~P.}\ \bibnamefont {Gentile~Fusillo}}, \bibinfo {author} {\bibfnamefont {B.~H.}\ \bibnamefont {Dunlap}}, \bibinfo {author} {\bibfnamefont {B.~T.}\ \bibnamefont {G{\"a}nsicke}}, \bibinfo {author} {\bibfnamefont {M.~A.}\ \bibnamefont {Hollands}}, \bibinfo {author} {\bibfnamefont {J.~J.}\ \bibnamefont {Hermes}}, \bibinfo {author} {\bibfnamefont {T.~R.}\ \bibnamefont {Marsh}}, \bibinfo {author} {\bibfnamefont {E.}~\bibnamefont {Cukanovaite}},\ and\ \bibinfo {author} {\bibfnamefont {T.}~\bibnamefont {Cunningham}},\ }\bibfield  {title} {\bibinfo {title} {Core crystallization and pile-up in the cooling sequence of evolving white dwarfs},\ }\href {https://doi.org/10.1038/s41586-018-0791-x} {\bibfield  {journal} {\bibinfo  {journal} {Nature}\ }\textbf {\bibinfo {volume} {565}},\ \bibinfo {pages} {202} (\bibinfo {year}
  {2019})}\BibitemShut {NoStop}%
\bibitem [{\citenamefont {Cheng}\ \emph {et~al.}(2019)\citenamefont {Cheng}, \citenamefont {Cummings},\ and\ \citenamefont {Menard}}]{Cheng2019ApJ}%
  \BibitemOpen
  \bibfield  {author} {\bibinfo {author} {\bibfnamefont {S.}~\bibnamefont {Cheng}}, \bibinfo {author} {\bibfnamefont {J.~D.}\ \bibnamefont {Cummings}},\ and\ \bibinfo {author} {\bibfnamefont {B.}~\bibnamefont {Menard}},\ }\bibfield  {title} {\bibinfo {title} {A cooling anomaly of high-mass white dwarfs},\ }\href {https://doi.org/10.3847/1538-4357/ab4989} {\bibfield  {journal} {\bibinfo  {journal} {Astrophysical Journal}\ }\textbf {\bibinfo {volume} {886}},\ \bibinfo {pages} {100} (\bibinfo {year} {2019})}\BibitemShut {NoStop}%
\bibitem [{\citenamefont {Blouin}\ \emph {et~al.}(2021)\citenamefont {Blouin}, \citenamefont {Daligault},\ and\ \citenamefont {Saumon}}]{Blouin2021ApJLNe22}%
  \BibitemOpen
  \bibfield  {author} {\bibinfo {author} {\bibfnamefont {S.}~\bibnamefont {Blouin}}, \bibinfo {author} {\bibfnamefont {J.}~\bibnamefont {Daligault}},\ and\ \bibinfo {author} {\bibfnamefont {D.}~\bibnamefont {Saumon}},\ }\bibfield  {title} {\bibinfo {title} {22ne phase separation as a solution to the ultramassive white dwarf cooling anomaly},\ }\href {https://doi.org/10.3847/2041-8213/abf14b} {\bibfield  {journal} {\bibinfo  {journal} {Astrophysical Journal Letters}\ }\textbf {\bibinfo {volume} {911}},\ \bibinfo {pages} {L5} (\bibinfo {year} {2021})}\BibitemShut {NoStop}%
\bibitem [{\citenamefont {Bedard}\ \emph {et~al.}(2024)\citenamefont {Bedard}, \citenamefont {Blouin},\ and\ \citenamefont {Cheng}}]{Bedard2024NatureBuoyant}%
  \BibitemOpen
  \bibfield  {author} {\bibinfo {author} {\bibfnamefont {A.}~\bibnamefont {Bedard}}, \bibinfo {author} {\bibfnamefont {S.}~\bibnamefont {Blouin}},\ and\ \bibinfo {author} {\bibfnamefont {S.}~\bibnamefont {Cheng}},\ }\bibfield  {title} {\bibinfo {title} {Buoyant crystals halt the cooling of white dwarf stars},\ }\href {https://doi.org/10.1038/s41586-024-07102-y} {\bibfield  {journal} {\bibinfo  {journal} {Nature}\ }\textbf {\bibinfo {volume} {627}},\ \bibinfo {pages} {286} (\bibinfo {year} {2024})}\BibitemShut {NoStop}%
\bibitem [{\citenamefont {Horowitz}\ \emph {et~al.}(2007)\citenamefont {Horowitz}, \citenamefont {Berry},\ and\ \citenamefont {Brown}}]{Horowitz2007PRE}%
  \BibitemOpen
  \bibfield  {author} {\bibinfo {author} {\bibfnamefont {C.~J.}\ \bibnamefont {Horowitz}}, \bibinfo {author} {\bibfnamefont {D.~K.}\ \bibnamefont {Berry}},\ and\ \bibinfo {author} {\bibfnamefont {E.~F.}\ \bibnamefont {Brown}},\ }\bibfield  {title} {\bibinfo {title} {Phase separation in the crust of accreting neutron stars},\ }\href {https://doi.org/10.1103/PhysRevE.75.066101} {\bibfield  {journal} {\bibinfo  {journal} {Physical Review E}\ }\textbf {\bibinfo {volume} {75}},\ \bibinfo {pages} {066101} (\bibinfo {year} {2007})}\BibitemShut {NoStop}%
\bibitem [{\citenamefont {Ichimaru}(1982)}]{Ichimaru1982}%
  \BibitemOpen
  \bibfield  {author} {\bibinfo {author} {\bibfnamefont {S.}~\bibnamefont {Ichimaru}},\ }\bibfield  {title} {\bibinfo {title} {Strongly coupled plasmas: High-density classical plasmas and degenerate electron liquids},\ }\href {https://doi.org/10.1103/RevModPhys.54.1017} {\bibfield  {journal} {\bibinfo  {journal} {Reviews of Modern Physics}\ }\textbf {\bibinfo {volume} {54}},\ \bibinfo {pages} {1017} (\bibinfo {year} {1982})}\BibitemShut {NoStop}%
\bibitem [{\citenamefont {Dubin}\ and\ \citenamefont {O'Neil}(1999)}]{DubinONeil1999}%
  \BibitemOpen
  \bibfield  {author} {\bibinfo {author} {\bibfnamefont {D.~H.~E.}\ \bibnamefont {Dubin}}\ and\ \bibinfo {author} {\bibfnamefont {T.~M.}\ \bibnamefont {O'Neil}},\ }\bibfield  {title} {\bibinfo {title} {Trapped nonneutral plasmas, liquids, and crystals (the thermal equilibrium states)},\ }\href {https://doi.org/10.1103/RevModPhys.71.87} {\bibfield  {journal} {\bibinfo  {journal} {Reviews of Modern Physics}\ }\textbf {\bibinfo {volume} {71}},\ \bibinfo {pages} {87} (\bibinfo {year} {1999})}\BibitemShut {NoStop}%
\bibitem [{\citenamefont {Drewsen}(2015)}]{Drewsen2015PhysicaB}%
  \BibitemOpen
  \bibfield  {author} {\bibinfo {author} {\bibfnamefont {M.}~\bibnamefont {Drewsen}},\ }\bibfield  {title} {\bibinfo {title} {Ion coulomb crystals},\ }\href {https://doi.org/10.1016/j.physb.2014.11.050} {\bibfield  {journal} {\bibinfo  {journal} {Physica B: Condensed Matter}\ }\textbf {\bibinfo {volume} {460}},\ \bibinfo {pages} {105} (\bibinfo {year} {2015})}\BibitemShut {NoStop}%
\bibitem [{\citenamefont {Kozhberov}(2024)}]{Kozhberov2024}%
  \BibitemOpen
  \bibfield  {author} {\bibinfo {author} {\bibfnamefont {A.~A.}\ \bibnamefont {Kozhberov}},\ }\bibfield  {title} {\bibinfo {title} {Electrostatic energy of solid binary ionic mixtures},\ }\href@noop {} {\bibfield  {journal} {\bibinfo  {journal} {Physical Review E}\ }\textbf {\bibinfo {volume} {110}},\ \bibinfo {pages} {045206} (\bibinfo {year} {2024})}\BibitemShut {NoStop}%
\bibitem [{\citenamefont {Tong}\ \emph {et~al.}(2010)\citenamefont {Tong}, \citenamefont {Winney},\ and\ \citenamefont {Willitsch}}]{PhysRevLettxx.105.143001}%
  \BibitemOpen
  \bibfield  {author} {\bibinfo {author} {\bibfnamefont {X.}~\bibnamefont {Tong}}, \bibinfo {author} {\bibfnamefont {A.~H.}\ \bibnamefont {Winney}},\ and\ \bibinfo {author} {\bibfnamefont {S.}~\bibnamefont {Willitsch}},\ }\bibfield  {title} {\bibinfo {title} {Sympathetic cooling of molecular ions in selected rotational and vibrational states produced by threshold photoionization},\ }\href {https://doi.org/10.1103/PhysRevLett.105.143001} {\bibfield  {journal} {\bibinfo  {journal} {Phys. Rev. Lett.}\ }\textbf {\bibinfo {volume} {105}},\ \bibinfo {pages} {143001} (\bibinfo {year} {2010})}\BibitemShut {NoStop}%
\bibitem [{\citenamefont {Chou}\ \emph {et~al.}(2017)\citenamefont {Chou}, \citenamefont {Kurz}, \citenamefont {Hume}, \citenamefont {Plessow}, \citenamefont {Leibrandt},\ and\ \citenamefont {Leibfried}}]{chou2017preparation}%
  \BibitemOpen
  \bibfield  {author} {\bibinfo {author} {\bibfnamefont {C.-w.}\ \bibnamefont {Chou}}, \bibinfo {author} {\bibfnamefont {C.}~\bibnamefont {Kurz}}, \bibinfo {author} {\bibfnamefont {D.~B.}\ \bibnamefont {Hume}}, \bibinfo {author} {\bibfnamefont {P.~N.}\ \bibnamefont {Plessow}}, \bibinfo {author} {\bibfnamefont {D.~R.}\ \bibnamefont {Leibrandt}},\ and\ \bibinfo {author} {\bibfnamefont {D.}~\bibnamefont {Leibfried}},\ }\bibfield  {title} {\bibinfo {title} {Preparation and coherent manipulation of pure quantum states of a single molecular ion},\ }\href {https://doi.org/https://doi.org/10.1038/nature22338} {\bibfield  {journal} {\bibinfo  {journal} {Nature}\ }\textbf {\bibinfo {volume} {545}},\ \bibinfo {pages} {203} (\bibinfo {year} {2017})}\BibitemShut {NoStop}%
\bibitem [{\citenamefont {Krohn}\ \emph {et~al.}(2023)\citenamefont {Krohn}, \citenamefont {Catani}, \citenamefont {P.~Sundar}, \citenamefont {Greenberg}, \citenamefont {da~Silva},\ and\ \citenamefont {Lewandowski}}]{krohn2023reactions}%
  \BibitemOpen
  \bibfield  {author} {\bibinfo {author} {\bibfnamefont {O.}~\bibnamefont {Krohn}}, \bibinfo {author} {\bibfnamefont {K.}~\bibnamefont {Catani}}, \bibinfo {author} {\bibfnamefont {S.}~\bibnamefont {P.~Sundar}}, \bibinfo {author} {\bibfnamefont {J.}~\bibnamefont {Greenberg}}, \bibinfo {author} {\bibfnamefont {G.}~\bibnamefont {da~Silva}},\ and\ \bibinfo {author} {\bibfnamefont {H.}~\bibnamefont {Lewandowski}},\ }\bibfield  {title} {\bibinfo {title} {Reactions of acetonitrile with trapped, translationally cold acetylene cations},\ }\href {https://doi.org/https://doi.org/10.1021/acs.jpca.3c00914} {\bibfield  {journal} {\bibinfo  {journal} {The Journal of Physical Chemistry A}\ }\textbf {\bibinfo {volume} {127}},\ \bibinfo {pages} {5120} (\bibinfo {year} {2023})}\BibitemShut {NoStop}%
\bibitem [{\citenamefont {Blackburn}\ and\ \citenamefont {Keller}(2020)}]{Blackburn2020SciRep}%
  \BibitemOpen
  \bibfield  {author} {\bibinfo {author} {\bibfnamefont {L.}~\bibnamefont {Blackburn}}\ and\ \bibinfo {author} {\bibfnamefont {M.}~\bibnamefont {Keller}},\ }\bibfield  {title} {\bibinfo {title} {The effect of the electric trapping field on state-selective loading of molecules into rf ion traps},\ }\href {https://doi.org/10.1038/s41598-020-74759-6} {\bibfield  {journal} {\bibinfo  {journal} {Scientific Reports}\ }\textbf {\bibinfo {volume} {10}},\ \bibinfo {pages} {18449} (\bibinfo {year} {2020})}\BibitemShut {NoStop}%
\bibitem [{\citenamefont {Blackburn}\ \emph {et~al.}(2025)\citenamefont {Blackburn}, \citenamefont {Shepherd},\ and\ \citenamefont {Keller}}]{Blackburn2025NJP}%
  \BibitemOpen
  \bibfield  {author} {\bibinfo {author} {\bibfnamefont {L.}~\bibnamefont {Blackburn}}, \bibinfo {author} {\bibfnamefont {A.}~\bibnamefont {Shepherd}},\ and\ \bibinfo {author} {\bibfnamefont {M.}~\bibnamefont {Keller}},\ }\bibfield  {title} {\bibinfo {title} {Fast switching of the rf trapping field in an ion trap},\ }\href {https://doi.org/10.1088/1367-2630/adc6b0} {\bibfield  {journal} {\bibinfo  {journal} {New Journal of Physics}\ }\textbf {\bibinfo {volume} {27}},\ \bibinfo {pages} {045001} (\bibinfo {year} {2025})}\BibitemShut {NoStop}%
\bibitem [{\citenamefont {Khanyile}\ \emph {et~al.}(2015)\citenamefont {Khanyile}, \citenamefont {Shu},\ and\ \citenamefont {Brown}}]{khanyile2015observation}%
  \BibitemOpen
  \bibfield  {author} {\bibinfo {author} {\bibfnamefont {N.~B.}\ \bibnamefont {Khanyile}}, \bibinfo {author} {\bibfnamefont {G.}~\bibnamefont {Shu}},\ and\ \bibinfo {author} {\bibfnamefont {K.~R.}\ \bibnamefont {Brown}},\ }\bibfield  {title} {\bibinfo {title} {Observation of vibrational overtones by single-molecule resonant photodissociation},\ }\href@noop {} {\bibfield  {journal} {\bibinfo  {journal} {Nature Communications}\ }\textbf {\bibinfo {volume} {6}},\ \bibinfo {pages} {7825} (\bibinfo {year} {2015})}\BibitemShut {NoStop}%
\bibitem [{\citenamefont {Rugango}\ \emph {et~al.}(2015)\citenamefont {Rugango}, \citenamefont {Goeders}, \citenamefont {Dixon}, \citenamefont {Gray}, \citenamefont {Khanyile}, \citenamefont {Shu}, \citenamefont {Clark},\ and\ \citenamefont {Brown}}]{rugango2015sympathetic}%
  \BibitemOpen
  \bibfield  {author} {\bibinfo {author} {\bibfnamefont {R.}~\bibnamefont {Rugango}}, \bibinfo {author} {\bibfnamefont {J.~E.}\ \bibnamefont {Goeders}}, \bibinfo {author} {\bibfnamefont {T.~H.}\ \bibnamefont {Dixon}}, \bibinfo {author} {\bibfnamefont {J.~M.}\ \bibnamefont {Gray}}, \bibinfo {author} {\bibfnamefont {N.}~\bibnamefont {Khanyile}}, \bibinfo {author} {\bibfnamefont {G.}~\bibnamefont {Shu}}, \bibinfo {author} {\bibfnamefont {R.~J.}\ \bibnamefont {Clark}},\ and\ \bibinfo {author} {\bibfnamefont {K.~R.}\ \bibnamefont {Brown}},\ }\bibfield  {title} {\bibinfo {title} {Sympathetic cooling of molecular ion motion to the ground state},\ }\href@noop {} {\bibfield  {journal} {\bibinfo  {journal} {New Journal of Physics}\ }\textbf {\bibinfo {volume} {17}},\ \bibinfo {pages} {035009} (\bibinfo {year} {2015})}\BibitemShut {NoStop}%
\bibitem [{\citenamefont {Okada}\ \emph {et~al.}(2015)\citenamefont {Okada}, \citenamefont {Ichikawa}, \citenamefont {Wada},\ and\ \citenamefont {Schuessler}}]{okPhysRevApplied.4.054009}%
  \BibitemOpen
  \bibfield  {author} {\bibinfo {author} {\bibfnamefont {K.}~\bibnamefont {Okada}}, \bibinfo {author} {\bibfnamefont {M.}~\bibnamefont {Ichikawa}}, \bibinfo {author} {\bibfnamefont {M.}~\bibnamefont {Wada}},\ and\ \bibinfo {author} {\bibfnamefont {H.~A.}\ \bibnamefont {Schuessler}},\ }\bibfield  {title} {\bibinfo {title} {Quasiequilibrium characterization of mixed-ion coulomb crystals},\ }\href {https://doi.org/10.1103/PhysRevApplied.4.054009} {\bibfield  {journal} {\bibinfo  {journal} {Phys. Rev. Appl.}\ }\textbf {\bibinfo {volume} {4}},\ \bibinfo {pages} {054009} (\bibinfo {year} {2015})}\BibitemShut {NoStop}%
\bibitem [{\citenamefont {Petralia}\ \emph {et~al.}(2020)\citenamefont {Petralia}, \citenamefont {Tsikritea}, \citenamefont {Loreau}, \citenamefont {Softley},\ and\ \citenamefont {Heazlewood}}]{petralia2020strong}%
  \BibitemOpen
  \bibfield  {author} {\bibinfo {author} {\bibfnamefont {L.}~\bibnamefont {Petralia}}, \bibinfo {author} {\bibfnamefont {A.}~\bibnamefont {Tsikritea}}, \bibinfo {author} {\bibfnamefont {J.}~\bibnamefont {Loreau}}, \bibinfo {author} {\bibfnamefont {T.}~\bibnamefont {Softley}},\ and\ \bibinfo {author} {\bibfnamefont {B.}~\bibnamefont {Heazlewood}},\ }\bibfield  {title} {\bibinfo {title} {Strong inverse kinetic isotope effect observed in ammonia charge exchange reactions},\ }\href@noop {} {\bibfield  {journal} {\bibinfo  {journal} {Nature Communications}\ }\textbf {\bibinfo {volume} {11}},\ \bibinfo {pages} {173} (\bibinfo {year} {2020})}\BibitemShut {NoStop}%
\bibitem [{\citenamefont {Guggemos}\ \emph {et~al.}(2015)\citenamefont {Guggemos}, \citenamefont {Heinrich}, \citenamefont {Herrera-Sancho}, \citenamefont {Blatt},\ and\ \citenamefont {Roos}}]{guggemos2015sympathetic}%
  \BibitemOpen
  \bibfield  {author} {\bibinfo {author} {\bibfnamefont {M.}~\bibnamefont {Guggemos}}, \bibinfo {author} {\bibfnamefont {D.}~\bibnamefont {Heinrich}}, \bibinfo {author} {\bibfnamefont {O.}~\bibnamefont {Herrera-Sancho}}, \bibinfo {author} {\bibfnamefont {R.}~\bibnamefont {Blatt}},\ and\ \bibinfo {author} {\bibfnamefont {C.}~\bibnamefont {Roos}},\ }\bibfield  {title} {\bibinfo {title} {Sympathetic cooling and detection of a hot trapped ion by a cold one},\ }\href@noop {} {\bibfield  {journal} {\bibinfo  {journal} {New Journal of Physics}\ }\textbf {\bibinfo {volume} {17}},\ \bibinfo {pages} {103001} (\bibinfo {year} {2015})}\BibitemShut {NoStop}%
\bibitem [{\citenamefont {Schmöger}\ \emph {et~al.}(2015)\citenamefont {Schmöger}, \citenamefont {Versolato}, \citenamefont {Schwarz}, \citenamefont {Kohnen}, \citenamefont {Windberger}, \citenamefont {Piest}, \citenamefont {Feuchtenbeiner}, \citenamefont {Pedregosa-Gutierrez}, \citenamefont {Leopold}, \citenamefont {Micke}, \citenamefont {Hansen}, \citenamefont {Baumann}, \citenamefont {Drewsen}, \citenamefont {Ullrich}, \citenamefont {Schmidt},\ and\ \citenamefont {López-Urrutia}}]{Schmoger2015Science}%
  \BibitemOpen
  \bibfield  {author} {\bibinfo {author} {\bibfnamefont {L.}~\bibnamefont {Schmöger}}, \bibinfo {author} {\bibfnamefont {O.~O.}\ \bibnamefont {Versolato}}, \bibinfo {author} {\bibfnamefont {M.}~\bibnamefont {Schwarz}}, \bibinfo {author} {\bibfnamefont {M.}~\bibnamefont {Kohnen}}, \bibinfo {author} {\bibfnamefont {A.}~\bibnamefont {Windberger}}, \bibinfo {author} {\bibfnamefont {B.}~\bibnamefont {Piest}}, \bibinfo {author} {\bibfnamefont {S.}~\bibnamefont {Feuchtenbeiner}}, \bibinfo {author} {\bibfnamefont {J.}~\bibnamefont {Pedregosa-Gutierrez}}, \bibinfo {author} {\bibfnamefont {T.}~\bibnamefont {Leopold}}, \bibinfo {author} {\bibfnamefont {P.}~\bibnamefont {Micke}}, \bibinfo {author} {\bibfnamefont {A.~K.}\ \bibnamefont {Hansen}}, \bibinfo {author} {\bibfnamefont {T.~M.}\ \bibnamefont {Baumann}}, \bibinfo {author} {\bibfnamefont {M.}~\bibnamefont {Drewsen}}, \bibinfo {author} {\bibfnamefont {J.}~\bibnamefont {Ullrich}}, \bibinfo {author} {\bibfnamefont {P.~O.}\ \bibnamefont {Schmidt}},\ and\ \bibinfo
  {author} {\bibfnamefont {J.~R.~C.}\ \bibnamefont {López-Urrutia}},\ }\bibfield  {title} {\bibinfo {title} {Coulomb crystallization of highly charged ions},\ }\href {https://doi.org/10.1126/science.aaa2960} {\bibfield  {journal} {\bibinfo  {journal} {Science}\ }\textbf {\bibinfo {volume} {347}},\ \bibinfo {pages} {1233} (\bibinfo {year} {2015})}\BibitemShut {NoStop}%
\bibitem [{\citenamefont {Chen}\ \emph {et~al.}(2025)\citenamefont {Chen}, \citenamefont {Zhou}, \citenamefont {Zhang}, \citenamefont {Xiao}, \citenamefont {Huang}, \citenamefont {Gao},\ and\ \citenamefont {Guan}}]{ChenShaolong2025}%
  \BibitemOpen
  \bibfield  {author} {\bibinfo {author} {\bibfnamefont {S.-L.}\ \bibnamefont {Chen}}, \bibinfo {author} {\bibfnamefont {Z.-q.}\ \bibnamefont {Zhou}}, \bibinfo {author} {\bibfnamefont {G.-s.}\ \bibnamefont {Zhang}}, \bibinfo {author} {\bibfnamefont {J.}~\bibnamefont {Xiao}}, \bibinfo {author} {\bibfnamefont {Y.}~\bibnamefont {Huang}}, \bibinfo {author} {\bibfnamefont {K.-L.}\ \bibnamefont {Gao}},\ and\ \bibinfo {author} {\bibfnamefont {H.}~\bibnamefont {Guan}},\ }\bibfield  {title} {\bibinfo {title} {Coulomb crystallization of highly charged ${\mathrm{ni}}^{12+}$ ions in a linear paul trap},\ }\href {https://doi.org/10.1103/cp1v-bnf1} {\bibfield  {journal} {\bibinfo  {journal} {Phys. Rev. A}\ }\textbf {\bibinfo {volume} {112}},\ \bibinfo {pages} {063115} (\bibinfo {year} {2025})}\BibitemShut {NoStop}%
\bibitem [{\citenamefont {Prokhorov}\ \emph {et~al.}(2025)\citenamefont {Prokhorov}, \citenamefont {Smith}, \citenamefont {Xu}, \citenamefont {Georgiou}, \citenamefont {Guarrera}, \citenamefont {Sajith}, \citenamefont {Dijck}, \citenamefont {Warnecke}, \citenamefont {Wehrheim}, \citenamefont {Wilzewski}, \citenamefont {Blackburn}, \citenamefont {Keller}, \citenamefont {Boyer}, \citenamefont {Pfeifer}, \citenamefont {Schwanke}, \citenamefont {Issever}, \citenamefont {Worm}, \citenamefont {Schmidt}, \citenamefont {Lopez-Urrutia},\ and\ \citenamefont {Barontini}}]{prokhorov2025coulombcrystallizationxenonhighly}%
  \BibitemOpen
  \bibfield  {author} {\bibinfo {author} {\bibfnamefont {L.}~\bibnamefont {Prokhorov}}, \bibinfo {author} {\bibfnamefont {A.~A.}\ \bibnamefont {Smith}}, \bibinfo {author} {\bibfnamefont {M.}~\bibnamefont {Xu}}, \bibinfo {author} {\bibfnamefont {K.}~\bibnamefont {Georgiou}}, \bibinfo {author} {\bibfnamefont {V.}~\bibnamefont {Guarrera}}, \bibinfo {author} {\bibfnamefont {L.~P.~K.}\ \bibnamefont {Sajith}}, \bibinfo {author} {\bibfnamefont {E.~A.}\ \bibnamefont {Dijck}}, \bibinfo {author} {\bibfnamefont {C.}~\bibnamefont {Warnecke}}, \bibinfo {author} {\bibfnamefont {M.}~\bibnamefont {Wehrheim}}, \bibinfo {author} {\bibfnamefont {A.}~\bibnamefont {Wilzewski}}, \bibinfo {author} {\bibfnamefont {L.}~\bibnamefont {Blackburn}}, \bibinfo {author} {\bibfnamefont {M.}~\bibnamefont {Keller}}, \bibinfo {author} {\bibfnamefont {V.}~\bibnamefont {Boyer}}, \bibinfo {author} {\bibfnamefont {T.}~\bibnamefont {Pfeifer}}, \bibinfo {author} {\bibfnamefont {U.}~\bibnamefont {Schwanke}}, \bibinfo {author} {\bibfnamefont
  {C.}~\bibnamefont {Issever}}, \bibinfo {author} {\bibfnamefont {S.}~\bibnamefont {Worm}}, \bibinfo {author} {\bibfnamefont {P.~O.}\ \bibnamefont {Schmidt}}, \bibinfo {author} {\bibfnamefont {J.~R.~C.}\ \bibnamefont {Lopez-Urrutia}},\ and\ \bibinfo {author} {\bibfnamefont {G.}~\bibnamefont {Barontini}},\ }\href {https://arxiv.org/abs/2512.12266} {\bibinfo {title} {Coulomb crystallization of xenon highly charged ions in a laser-cooled ca+ matrix}} (\bibinfo {year} {2025}),\ \Eprint {https://arxiv.org/abs/2512.12266} {arXiv:2512.12266 [physics.atom-ph]} \BibitemShut {NoStop}%
\bibitem [{SM()}]{SM}%
  \BibitemOpen
  \href@noop {} {}\bibinfo {note} {See Supplemental Material for details, which include \cite{Frohlich2005,Okada2010,d2015simple,Schlag1983Tenengrad,Pertuz2013FocusMeasures,Khrapak2015}}\BibitemShut {NoStop}%
\bibitem [{\citenamefont {R\"uffert}\ \emph {et~al.}(2024)\citenamefont {R\"uffert}, \citenamefont {Dijck}, \citenamefont {Timm}, \citenamefont {{Crespo L\'opez-Urrutia}},\ and\ \citenamefont {Mehlst\"aubler}}]{Rueffert2024}%
  \BibitemOpen
  \bibfield  {author} {\bibinfo {author} {\bibfnamefont {L.-A.}\ \bibnamefont {R\"uffert}}, \bibinfo {author} {\bibfnamefont {E.~A.}\ \bibnamefont {Dijck}}, \bibinfo {author} {\bibfnamefont {L.}~\bibnamefont {Timm}}, \bibinfo {author} {\bibfnamefont {J.~R.}\ \bibnamefont {{Crespo L\'opez-Urrutia}}},\ and\ \bibinfo {author} {\bibfnamefont {T.~E.}\ \bibnamefont {Mehlst\"aubler}},\ }\bibfield  {title} {\bibinfo {title} {Domain formation and structural stabilities in mixed-species coulomb crystals induced by sympathetically cooled highly charged ions},\ }\href {https://doi.org/10.1103/PhysRevA.110.063110} {\bibfield  {journal} {\bibinfo  {journal} {Phys. Rev. A}\ }\textbf {\bibinfo {volume} {110}},\ \bibinfo {pages} {063110} (\bibinfo {year} {2024})}\BibitemShut {NoStop}%
\bibitem [{\citenamefont {Mestel}(1952)}]{Mestel1952}%
  \BibitemOpen
  \bibfield  {author} {\bibinfo {author} {\bibfnamefont {L.}~\bibnamefont {Mestel}},\ }\bibfield  {title} {\bibinfo {title} {On the theory of white dwarf stars: I. the energy sources of white dwarfs},\ }\href {https://doi.org/10.1093/mnras/112.6.583} {\bibfield  {journal} {\bibinfo  {journal} {Monthly Notices of the Royal Astronomical Society}\ }\textbf {\bibinfo {volume} {112}},\ \bibinfo {pages} {583} (\bibinfo {year} {1952})}\BibitemShut {NoStop}%
\bibitem [{\citenamefont {Shapiro}\ and\ \citenamefont {Teukolsky}(1983)}]{ShapiroTeukolsky}%
  \BibitemOpen
  \bibfield  {author} {\bibinfo {author} {\bibfnamefont {S.~L.}\ \bibnamefont {Shapiro}}\ and\ \bibinfo {author} {\bibfnamefont {S.~A.}\ \bibnamefont {Teukolsky}},\ }\href {https://doi.org/10.1002/9783527617661} {\emph {\bibinfo {title} {Black Holes, White Dwarfs, and Neutron Stars: The Physics of Compact Objects}}}\ (\bibinfo  {publisher} {Wiley},\ \bibinfo {address} {New York},\ \bibinfo {year} {1983})\BibitemShut {NoStop}%
\bibitem [{\citenamefont {Bhattacharya}\ \emph {et~al.}(2018)\citenamefont {Bhattacharya}, \citenamefont {Mukhopadhyay},\ and\ \citenamefont {Mukerjee}}]{Bhattacharya2018MNRAS_Mestel}%
  \BibitemOpen
  \bibfield  {author} {\bibinfo {author} {\bibfnamefont {M.}~\bibnamefont {Bhattacharya}}, \bibinfo {author} {\bibfnamefont {B.}~\bibnamefont {Mukhopadhyay}},\ and\ \bibinfo {author} {\bibfnamefont {S.}~\bibnamefont {Mukerjee}},\ }\bibfield  {title} {\bibinfo {title} {Luminosity and cooling of highly magnetized white dwarfs: suppression of luminosity by strong magnetic fields},\ }\href {https://doi.org/10.1093/mnras/sty776} {\bibfield  {journal} {\bibinfo  {journal} {Monthly Notices of the Royal Astronomical Society}\ }\textbf {\bibinfo {volume} {477}},\ \bibinfo {pages} {2705} (\bibinfo {year} {2018})}\BibitemShut {NoStop}%
\bibitem [{\citenamefont {Haensel}\ \emph {et~al.}(2007)\citenamefont {Haensel}, \citenamefont {Potekhin},\ and\ \citenamefont {Yakovlev}}]{HaenselPotekhinYakovlev2007}%
  \BibitemOpen
  \bibfield  {author} {\bibinfo {author} {\bibfnamefont {P.}~\bibnamefont {Haensel}}, \bibinfo {author} {\bibfnamefont {A.~Y.}\ \bibnamefont {Potekhin}},\ and\ \bibinfo {author} {\bibfnamefont {D.~G.}\ \bibnamefont {Yakovlev}},\ }\href {https://doi.org/10.1007/978-0-387-47301-7} {\emph {\bibinfo {title} {Neutron Stars 1: Equation of State and Structure}}},\ \bibinfo {series} {Astrophysics and Space Science Library}, Vol.\ \bibinfo {volume} {326}\ (\bibinfo  {publisher} {Springer},\ \bibinfo {year} {2007})\BibitemShut {NoStop}%
\bibitem [{\citenamefont {Hamaguchi}\ \emph {et~al.}(1997)\citenamefont {Hamaguchi}, \citenamefont {Farouki},\ and\ \citenamefont {Dubin}}]{Hamaguchi1997}%
  \BibitemOpen
  \bibfield  {author} {\bibinfo {author} {\bibfnamefont {S.}~\bibnamefont {Hamaguchi}}, \bibinfo {author} {\bibfnamefont {R.~T.}\ \bibnamefont {Farouki}},\ and\ \bibinfo {author} {\bibfnamefont {D.~H.~E.}\ \bibnamefont {Dubin}},\ }\bibfield  {title} {\bibinfo {title} {Triple point of yukawa systems},\ }\href {https://doi.org/10.1103/PhysRevE.56.4671} {\bibfield  {journal} {\bibinfo  {journal} {Physical Review E}\ }\textbf {\bibinfo {volume} {56}},\ \bibinfo {pages} {4671} (\bibinfo {year} {1997})}\BibitemShut {NoStop}%
\bibitem [{\citenamefont {Stojanovi{\'c}}\ \emph {et~al.}(2012)\citenamefont {Stojanovi{\'c}}, \citenamefont {Shi}, \citenamefont {Bruder},\ and\ \citenamefont {Cirac}}]{Stojanovic2012PRL}%
  \BibitemOpen
  \bibfield  {author} {\bibinfo {author} {\bibfnamefont {V.~M.}\ \bibnamefont {Stojanovi{\'c}}}, \bibinfo {author} {\bibfnamefont {T.}~\bibnamefont {Shi}}, \bibinfo {author} {\bibfnamefont {C.}~\bibnamefont {Bruder}},\ and\ \bibinfo {author} {\bibfnamefont {J.~I.}\ \bibnamefont {Cirac}},\ }\bibfield  {title} {\bibinfo {title} {Quantum simulation of small-polaron formation with trapped ions},\ }\href {https://doi.org/10.1103/PhysRevLett.109.250501} {\bibfield  {journal} {\bibinfo  {journal} {Physical Review Letters}\ }\textbf {\bibinfo {volume} {109}},\ \bibinfo {pages} {250501} (\bibinfo {year} {2012})}\BibitemShut {NoStop}%
\bibitem [{\citenamefont {Fr{\"o}hlich}\ \emph {et~al.}(2005)\citenamefont {Fr{\"o}hlich}, \citenamefont {Roth},\ and\ \citenamefont {Schiller}}]{Frohlich2005}%
  \BibitemOpen
  \bibfield  {author} {\bibinfo {author} {\bibfnamefont {U.}~\bibnamefont {Fr{\"o}hlich}}, \bibinfo {author} {\bibfnamefont {B.}~\bibnamefont {Roth}},\ and\ \bibinfo {author} {\bibfnamefont {S.}~\bibnamefont {Schiller}},\ }\bibfield  {title} {\bibinfo {title} {Ellipsoidal coulomb crystals in a linear radio-frequency trap},\ }\href {https://doi.org/10.1063/1.1976605} {\bibfield  {journal} {\bibinfo  {journal} {Physics of Plasmas}\ }\textbf {\bibinfo {volume} {12}},\ \bibinfo {pages} {073506} (\bibinfo {year} {2005})}\BibitemShut {NoStop}%
\bibitem [{\citenamefont {Okada}\ \emph {et~al.}(2010)\citenamefont {Okada}, \citenamefont {Wada}, \citenamefont {Takayanagi}, \citenamefont {Ohtani},\ and\ \citenamefont {Schuessler}}]{Okada2010}%
  \BibitemOpen
  \bibfield  {author} {\bibinfo {author} {\bibfnamefont {K.}~\bibnamefont {Okada}}, \bibinfo {author} {\bibfnamefont {M.}~\bibnamefont {Wada}}, \bibinfo {author} {\bibfnamefont {T.}~\bibnamefont {Takayanagi}}, \bibinfo {author} {\bibfnamefont {S.}~\bibnamefont {Ohtani}},\ and\ \bibinfo {author} {\bibfnamefont {H.~A.}\ \bibnamefont {Schuessler}},\ }\bibfield  {title} {\bibinfo {title} {Characterization of ion coulomb crystals in a linear paul trap},\ }\href {https://doi.org/10.1103/PhysRevA.81.013420} {\bibfield  {journal} {\bibinfo  {journal} {Physical Review A}\ }\textbf {\bibinfo {volume} {81}},\ \bibinfo {pages} {013420} (\bibinfo {year} {2010})}\BibitemShut {NoStop}%
\bibitem [{\citenamefont {D’yachkov}(2015)}]{d2015simple}%
  \BibitemOpen
  \bibfield  {author} {\bibinfo {author} {\bibfnamefont {L.~G.}\ \bibnamefont {D’yachkov}},\ }\bibfield  {title} {\bibinfo {title} {A simple analytical model of the coulomb cluster in a cylindrically symmetric parabolic trap},\ }\href {https://doi.org/10.1134/S0018151X15050107} {\bibfield  {journal} {\bibinfo  {journal} {High Temperature}\ }\textbf {\bibinfo {volume} {53}},\ \bibinfo {pages} {613} (\bibinfo {year} {2015})}\BibitemShut {NoStop}%
\bibitem [{\citenamefont {Schlag}\ \emph {et~al.}(1983)\citenamefont {Schlag}, \citenamefont {Sanderson}, \citenamefont {Neuman},\ and\ \citenamefont {Wimberly}}]{Schlag1983Tenengrad}%
  \BibitemOpen
  \bibfield  {author} {\bibinfo {author} {\bibfnamefont {J.~F.}\ \bibnamefont {Schlag}}, \bibinfo {author} {\bibfnamefont {A.~C.}\ \bibnamefont {Sanderson}}, \bibinfo {author} {\bibfnamefont {C.~P.}\ \bibnamefont {Neuman}},\ and\ \bibinfo {author} {\bibfnamefont {F.~C.}\ \bibnamefont {Wimberly}},\ }\href@noop {} {\emph {\bibinfo {title} {Implementation of Automatic Focusing Algorithms for a Computer Vision System with Camera Control}}},\ \bibinfo {type} {Tech. Rep.}\ \bibinfo {number} {CMU-RI-TR-83-14}\ (\bibinfo  {institution} {Robotics Institute, Carnegie Mellon University},\ \bibinfo {year} {1983})\BibitemShut {NoStop}%
\bibitem [{\citenamefont {Pertuz}\ \emph {et~al.}(2013)\citenamefont {Pertuz}, \citenamefont {Puig},\ and\ \citenamefont {Garcia}}]{Pertuz2013FocusMeasures}%
  \BibitemOpen
  \bibfield  {author} {\bibinfo {author} {\bibfnamefont {S.}~\bibnamefont {Pertuz}}, \bibinfo {author} {\bibfnamefont {D.}~\bibnamefont {Puig}},\ and\ \bibinfo {author} {\bibfnamefont {M.~A.}\ \bibnamefont {Garcia}},\ }\bibfield  {title} {\bibinfo {title} {Analysis of focus measure operators for shape-from-focus},\ }\href {https://doi.org/10.1016/j.patcog.2012.11.011} {\bibfield  {journal} {\bibinfo  {journal} {Pattern Recognition}\ }\textbf {\bibinfo {volume} {46}},\ \bibinfo {pages} {1415} (\bibinfo {year} {2013})}\BibitemShut {NoStop}%
\bibitem [{\citenamefont {Khrapak}\ \emph {et~al.}(2015)\citenamefont {Khrapak}, \citenamefont {Kryuchkov}, \citenamefont {Yurchenko},\ and\ \citenamefont {Thomas}}]{Khrapak2015}%
  \BibitemOpen
  \bibfield  {author} {\bibinfo {author} {\bibfnamefont {S.~A.}\ \bibnamefont {Khrapak}}, \bibinfo {author} {\bibfnamefont {N.~P.}\ \bibnamefont {Kryuchkov}}, \bibinfo {author} {\bibfnamefont {S.~O.}\ \bibnamefont {Yurchenko}},\ and\ \bibinfo {author} {\bibfnamefont {H.~M.}\ \bibnamefont {Thomas}},\ }\bibfield  {title} {\bibinfo {title} {Practical thermodynamics of yukawa systems at strong coupling},\ }\href {https://doi.org/10.1063/1.4921223} {\bibfield  {journal} {\bibinfo  {journal} {The Journal of Chemical Physics}\ }\textbf {\bibinfo {volume} {142}},\ \bibinfo {pages} {194903} (\bibinfo {year} {2015})}\BibitemShut {NoStop}%
\end{thebibliography}%

\newpage

\clearpage
\onecolumngrid

\renewcommand{\thefigure}{S\arabic{figure}}
\setcounter{figure}{0}
\section*{Supplementary Materials}

\subsection*{Ion number and impurity parameter}

For the large crystals, where individual ions are not resolved in the fluorescence images, we estimate the Ca$^+$ host number $N_{ Ca}$ from the undoped reference crystal recorded for a specified confinement setting. In the cold fluid description appropriate for a linear Paul trap, the crystal is treated as a uniform-density ellipsoid. The equilibrium density is fixed by the trap pseudopotential and, in terms of the secular frequencies, is given by \cite{Frohlich2005,Okada2010,DubinONeil1999}
\begin{equation}
n=\frac{\varepsilon_0 m_{ Ca}}{e^2}\left(\omega_x^2+\omega_y^2+\omega_z^2\right).
\end{equation}
The corresponding bulk estimate of the ion number is then obtained from the ellipsoidal volume,
\begin{equation}
N_{ Ca}=n\frac{4\pi}{3}R_xR_yR_z.
\end{equation}
Here $R_z$ is obtained from the measured axial extent, whereas only one radial semiaxis is directly visible in the image. We therefore consider the two possible assignments of the measured radial size to the transverse trap axes and use the frequency ratio $\omega_x/\omega_y$ to infer the unseen semiaxis. This yields two atom-number estimates, $N_1$ and $N_2$, from which we define
\begin{equation}
N_{Ca}=\frac{N_1+N_2}{2},
\qquad
\delta N_{Ca}=\frac{|N_1-N_2|}{2}.
\end{equation}

For smaller crystals, the continuum bulk estimate must be corrected for finite size effects. We utilize the correction developed in \cite{d2015simple}, according to which the linear cluster size is reduced relative to the large $N$ homogeneous limit by the factor
\begin{equation}
f(N)=1-0.9\left(\frac{4}{3\pi N}\right)^{1/3}.
\end{equation}
Hence we determine $N_{ Ca}$ from the implicit relation
\begin{equation}
V_{ meas}=\frac{N_{ Ca}}{n}f(N_{ Ca})^3,
\end{equation}
where $V_{ meas}=(4\pi/3)R_xR_yR_z$ is the volume reconstructed from the measured semiaxes. 

The trapping frequencies are measured independently on crystals containing only a handful of Ca$^+$ and by modulating the
amplitude of the RF drive of the Paul trap until some axial or radial motion is excited.  The number of Xe$^{12+}$ ions $N_{Xe}$ is estimated by simply counting the voids in the fluorescence images.

We quantify the impurity content via
\begin{equation}
Q_{imp}\equiv \langle Z^2\rangle-\langle Z\rangle^2,
\end{equation}
where the averages are taken over the ionic charge distribution. In our experiments the host is Ca$^+$ with $Z_{ Ca}=1$ and the dopant is Xe$^{12+}$ with $Z_{ Xe}=12$. For a two-component system with $N_{ Ca}$ host ions and $N_{ Xe}$ dopants,
\begin{equation}
\langle Z\rangle=\frac{N_{ Ca}Z_{ Ca}+N_{ Xe}Z_{ Xe}}{N_{ Ca}+N_{ Xe}},\qquad
\langle Z^2\rangle=\frac{N_{ Ca}Z_{ Ca}^2+N_{ Xe}Z_{ Xe}^2}{N_{ Ca}+N_{ Xe}}.
\end{equation}
Defining $r\equiv N_{ Xe}/N_{ Ca}$, we obtain:
\begin{equation}
Q_{imp}=\frac{121\,r}{(1+r)^2}.
\end{equation}

\subsection*{Evaluation of the Coulomb coupling parameter}

We characterize the Ca$^+$ host by the Coulomb coupling parameter
\begin{equation}
\Gamma \equiv \frac{Z_{ Ca}^2 e^2}{4\pi\varepsilon_0\, a\, k_B T},
\end{equation}
where $T$ is the Ca$^+$ temperature, and $a=[3/(4\pi n)]^{1/3}$ is the Wigner-Seitz radius of the Ca$^+$ host component. In our experiment the cooling laser is red-detuned by $\simeq 25~\mathrm{MHz}$, resulting in $T\simeq 0.7~\mathrm{mK}$. Changing the confinement can in principle modify residual RF-driven heating (e.g. via micromotion), introducing a systematic shift in $T$, that is difficult to quantify in large 3D crystals. Note however that such possible temperature related offsets are strongly suppressed in the ratio $\gamma=\Gamma_c(Q_{ imp})/\Gamma_c(0)$, which is our primary observable. 

We evaluate $n$ (and hence $a$) from the measured secular frequencies using the undoped Ca$^+$ reference, and we use this same $a$ to define the Ca$^+$ coupling parameter $\Gamma$ for all impurity settings at fixed confinement. In this way, $\Gamma$ is always referenced to the host crystal that would occupy the same trap in the absence of dopants. The Xe$^{12+}$ impurities then appear experimentally as a shift of the critical value $\Gamma_c$ on this fixed reference scale. This means that the mixed crystal is not re-parametrized in terms of a new effective density; rather, the impurity effect is quantified through the displacement of the crystallization threshold relative to the undoped host reference. The validity of this approximation is supported by the absence of a systematic dependence of the measured crystal radii on the number of implanted dopants within experimental resolution, and by the spatially localized character of the impurity-induced perturbation.

In the experiment we drive the liquid-to-crystal crossover by increasing the radial confinement. We use seven distinct confinement settings, corresponding to the Coulomb-parameter values $\Gamma_i$, computed as detailed above.

\subsection*{Relation between $\Gamma$ and mixed-species conventions}

It is worth noticing that there is no unique or universally adopted convention for defining a single Coulomb coupling parameter in a mixed-species plasma. In this work we therefore use the single-component Ca$^+$ $\Gamma$ as the control parameter, since it is directly anchored to experimentally accessible observables. Indeed, in our experiment the Xe$^{12+}$ ions cannot be directly imaged. Our choice of $\Gamma$ can nonetheless be related to standard mixed-species conventions. For example, in a binary Coulomb mixture one commonly introduces the electron-sphere radius
\begin{equation}
a_e=\left(\frac{3}{4\pi n_e}\right)^{1/3},
\end{equation}
with $n_e$ the electron density, and defines a species-resolved coupling parameter
\begin{equation}
\Gamma_j^{ (mix)}=\frac{Z_j^{5/3}e^2}{4\pi\varepsilon_0 a_e k_B T}.
\end{equation}
For the present Ca$^+$-Xe$^{12+}$ system, assuming, as discussed above, that the average Ca$^+$ host density is unchanged by the presence of a small number of dopants, one has
\begin{equation}
n_e=n_{ Ca}+12\,n_{ Xe}=n_{ Ca}(1+12r).
\end{equation}
It then follows that
\begin{equation}
\Gamma_{ Ca}^{ (mix)}=(1+12r)^{1/3}\Gamma,
\end{equation}
and therefore
\begin{equation}
\Gamma_{ Xe}^{ (mix)}=12^{5/3}\Gamma_{ Ca}^{ (mix)}.
\end{equation}
Similarly, by introducing a composition-averaged mixture parameter,
\begin{equation}
\Gamma_{ mix}=\langle Z^{5/3}\rangle\frac{e^2}{4\pi\varepsilon_0 a_e k_B T},
\end{equation}
one obtains
\begin{equation}
\Gamma_{ mix}
=
\frac{1+12^{5/3}r}{1+r}\,(1+12r)^{1/3}\,\Gamma .
\end{equation}

\subsection*{Tenengrad crystallinity metric}

For large Ca$^+$ crystals, single ions are not resolved and the Lindemann parameter cannot be extracted reliably from fluorescence images. We therefore quantify the emergence of crystalline order using an image-based sharpness metric, the Tenengrad (squared-gradient) measure, see e.g. \cite{Schlag1983Tenengrad,Pertuz2013FocusMeasures} . The key idea is that crystallization produces sharper spatial features in the fluorescence (higher spatial-frequency content), while the liquid phase appears smoother on the same pixel scale.

Starting from a fluorescence image $I(x,y)$, we compute the Sobel derivatives $g_x$ and $g_y$, i.e. discrete estimates of the intensity gradients obtained by convolving $I$ with the standard $3\times 3$ Sobel kernels
\begin{equation}
S_x=\begin{pmatrix}
-1 & 0 & 1\\
-2 & 0 & 2\\
-1 & 0 & 1
\end{pmatrix},\qquad
S_y=\begin{pmatrix}
-1 & -2 & -1\\
0 & 0 & 0\\
1 & 2 & 1
\end{pmatrix},
\end{equation}
so that $g_x = S_x * I$ and $g_y = S_y * I$, where $*$ denotes discrete convolution. We then form the Tenengrad energy map
\begin{equation}
\mathcal{T}(x,y)=g_x(x,y)^2+g_y(x,y)^2.
\end{equation}
For each frame, we compute the scalar Tenengrad value by averaging $\mathcal{T}(x,y)$ over the sample region,
\begin{equation}
\langle \mathcal{T}\rangle = \langle \mathcal{T}(x,y)\rangle_{ sample},
\end{equation}
where the sample region is selected to exclude background pixels. Indeed, to restrict the analysis to the fluorescing Ca$^+$ cloud, we apply an intensity mask. For each frame we compute a smoothed version of the image and determine a global threshold using Otsu's method; pixels above threshold define the sample region. We then compute $\langle\mathcal{T}\rangle$ by averaging $\mathcal{T}(x,y)$ over the masked region. We verified that $\langle\mathcal{T}\rangle$ and the extracted $\Gamma_c$ are insensitive, within scatter, to reasonable variations of the masking procedure (threshold choice and smoothing strength). By construction, $\langle\mathcal{T}\rangle$ provides a scalar measure of image sharpness that increases with the degree of spatial ordering of the Ca$^+$ fluorescence, and we therefore use it as a measure of the crystallinity of the sample. For each experimental setting we acquire five fluorescence frames and evaluate the mean and standard deviation of $\langle\mathcal{T}\rangle$.

It is important to note that $\mathcal{T}(x,y)$ is not an absolute measure of crystallinity. Because it is quadratic in the image gradients, $\langle\mathcal{T}\rangle$ is sensitive not only to spatial ordering but also to the overall fluorescence level: if the image intensity is globally rescaled by a factor $A$, one expects $g_x,g_y\propto A$ and therefore $\langle\mathcal{T}\rangle\propto A^2$. As a result, brighter clouds, i.e. clouds with more atoms, exhibit systematically larger $\langle\mathcal{T}\rangle$ even at comparable structural order. This must be accounted for in the extraction of $\Gamma_c$, as explained below.

\subsection*{Connection between Tenengrad sharpness and the Lindemann parameter}

As mentioned earlier, because a large fraction of the ions in large 3D crystals lie outside the focal plane, the resulting depth-of-field blur makes an absolute extraction of the Lindemann parameter from fluorescence images highly model-dependent. In contrast, the Tenengrad sharpness remains a well-defined, reproducible scalar that tracks the overall degree of ordering through the emergence of sharp fluorescence features. In this section we show, within a minimal imaging model, that the Tenengrad sharpness is monotonically related to the Lindemann parameter.

We model the fluorescence image as a sum of identical single-ion point-spread functions (PSFs) centered at the instantaneous ion positions. We assume that, during the camera exposure, each ion undergoes Gaussian fluctuations about its equilibrium site with in-plane rms displacement $u_{ rms}$. Approximating both the optical PSF and the motional blurring as Gaussians, the time-averaged image is equivalent to a static-lattice image convolved with a Gaussian kernel of effective width
\begin{equation}
\sigma_{ eff}^2=\sigma_0^2+u_{ rms}^2,
\end{equation}
where $\sigma_0$ is the (in-plane) optical PSF width. Identifying $u_{ rms}=\mathcal{L}\,a$ yields
\begin{equation}
\sigma_{ eff}^2=\sigma_0^2+\mathcal{L}^2 a^2.
\end{equation}

In the continuous limit the Tenengrad scalar is the squared-gradient energy of the image,
\begin{equation}
\langle\mathcal{T}\rangle \;\propto\; \int d^2r\,|\nabla I(\mathbf{r})|^2.
\end{equation}
Our $\langle\mathcal{T}\rangle$ is only proportional to this integral because it is a discretized, Sobel-kernel implementation of the squared-gradient functional. Therefore it matches the integral only up to an overall prefactor set by pixel size, kernel normalization, and imaging gain/exposure. Using Parseval's identity, this can be written in Fourier space as
\begin{equation}
\langle\mathcal{T}\rangle \;\propto\; \int d^2k\,k^2\,|I(\mathbf{k})|^2.
\end{equation}
Gaussian blurring multiplies the image spectrum by
\begin{equation}
I(\mathbf{k}) \rightarrow I(\mathbf{k})\,\exp\!\left(-\frac{\sigma_{ eff}^2 k^2}{2}\right),
\end{equation}
which suppresses high-$k$ components and therefore reduces $\langle\mathcal{T}\rangle$ as $\sigma_{ eff}$ increases.

Using a single-scale approximation in which the dominant spatial structure contributing to $\langle\mathcal{T}\rangle$ is characterized by a typical wave number $k_0=O(1/a)$, Gaussian blurring suppresses the corresponding Fourier components as $\exp(-\sigma_{ eff}^2 k_0^2)$. This gives
\begin{equation}
k_0=\frac{\alpha}{a}\quad (\alpha=O(1))
\;\;\Rightarrow\;\;
\langle\mathcal{T}\rangle \propto 
\exp\!\left[-\alpha^2\left(\frac{\sigma_0}{a}\right)^2\right]\,
\exp\!\left(-\alpha^2\mathcal{L}^2\right)
\equiv \langle\mathcal{T}\rangle_0\,\exp\!\left(-\alpha^2\mathcal{L}^2\right).
\end{equation}
For $\mathcal{L}\ll 1$,
\begin{equation}
\frac{\langle\mathcal{T}\rangle}{\langle\mathcal{T}\rangle_0}
=1-\alpha^2\mathcal{L}^2+\frac{\alpha^4}{2}\mathcal{L}^4+O(\mathcal{L}^6).
\end{equation}
which is strictly decreasing with $\mathcal{L}^2$. Therefore increasing ionic delocalization (larger $\mathcal{L}$) reduces the image sharpness and lowers $\langle\mathcal{T}\rangle$, providing a mathematical basis for the use of Tenengrad as a measure of crystallinity. Note that both $k_0$ and $\sigma_0$ depend on the imaging setup.

\subsection*{Extraction of $\Gamma_c$}
\label{sec:GammacExtraction}

In our inhomogeneous samples the liquid-to-solid transition is observed as a smooth crossover rather than a sharp discontinuity. We therefore extract an \emph{effective} crossover point $\Gamma_c$ from the measured Tenengrad sharpness curve $\langle\mathcal{T}\rangle(\Gamma)$ using a threshold-crossing procedure.

As discussed above, the absolute value of $\langle\mathcal{T}\rangle$ depends not only on structural order but also on the overall fluorescence level, and therefore on the number of Ca$^+$ ions in the sample (see for example Fig. S1). For this reason, we do not define a single global threshold for all datasets. Instead, for each Ca$^+$ host size (fixed $N_{ Ca}$) and each $N_{ Xe}$, we reconstruct the discrete curve $\langle\mathcal{T}\rangle(\Gamma_i)$. For each dataset we define a \emph{single} crystallinity threshold from the \emph{undoped} reference (no Xe$^{12+}$), and apply the same threshold to all doped samples prepared with the same host size. Specifically, denoting the undoped mean sharpness at $\Gamma_i$ by $\langle\mathcal{T}\rangle_0(\Gamma_i)$, we take
\begin{equation}
\langle\mathcal{T}\rangle_{ low}\equiv \langle\mathcal{T}\rangle_0(\Gamma_1),
\qquad
\langle\mathcal{T}\rangle_{ high}\equiv { median}\!\left\{\langle\mathcal{T}\rangle_0(\Gamma_5),\langle\mathcal{T}\rangle_0(\Gamma_6),\langle\mathcal{T}\rangle_0(\Gamma_7)\right\},
\end{equation}
and define the threshold as the midpoint
\begin{equation}
\langle\mathcal{T}\rangle_{ thr}\equiv \frac{\langle\mathcal{T}\rangle_{ low}+\langle\mathcal{T}\rangle_{ high}}{2}.
\end{equation}
This threshold definition is an operational choice for a smooth crossover, rather than a formally defined thermodynamic criterion. It therefore affects the absolute value assigned to $\Gamma_c$, in other words, it inevitably introduces a bias. However, for a given host size the same threshold is applied to all impurity settings; the threshold choice therefore shifts $\Gamma_c$ for doped and undoped data in the same way, and cancels in the normalized ratio $\gamma(Q_{imp})=\Gamma_c(Q_{imp})/\Gamma_c(0)$, which is the quantity reported in the main text.

\begin{figure}
\centering
\includegraphics[width=0.475\textwidth]{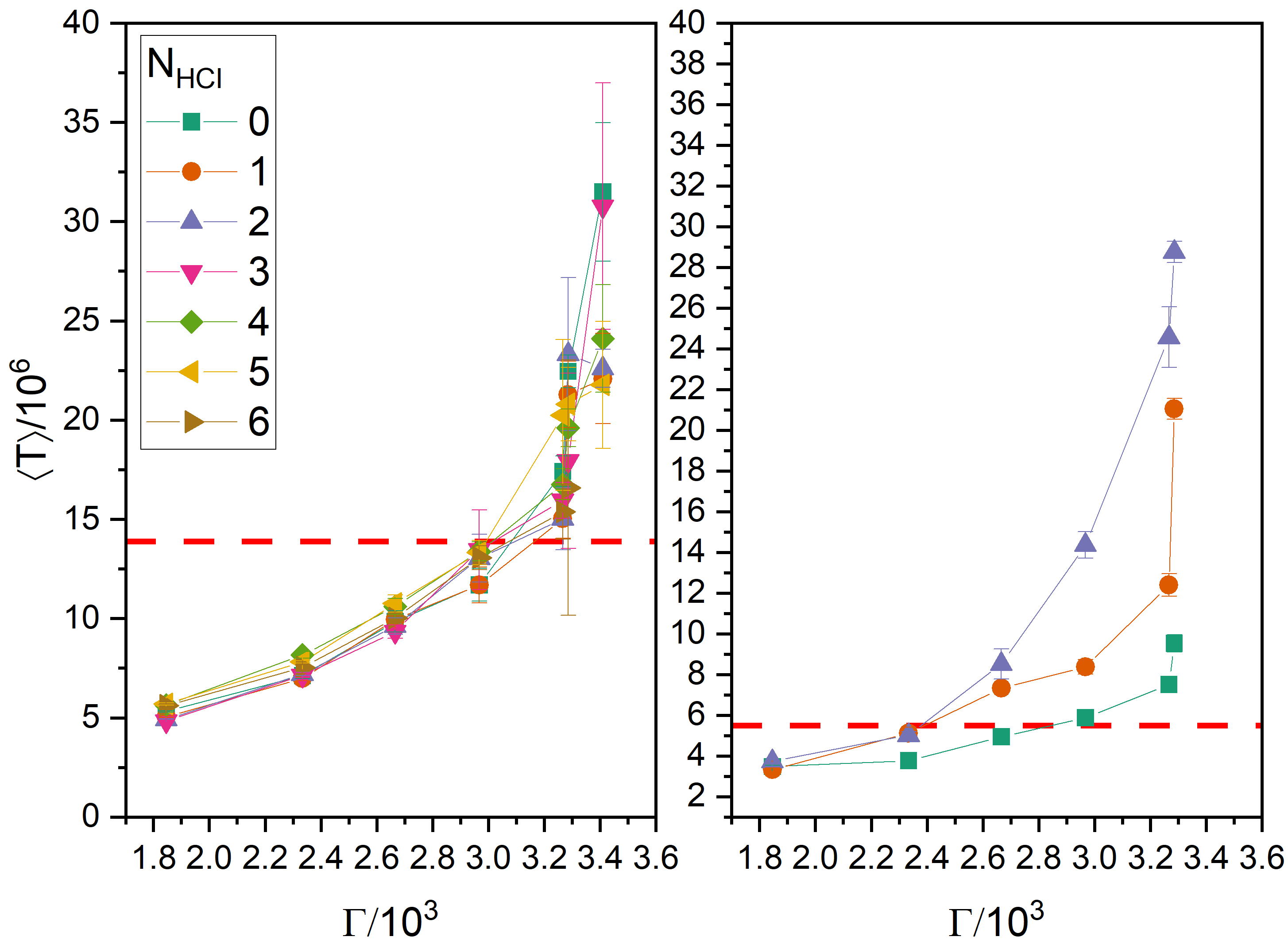}
\caption{Average Tenengrad $\langle\mathcal{T}\rangle$ for a sample of 250 $\pm$ 30 (left) and 21 $\pm$ 2 (right) Ca$^+$ ions containing a variable number of Xe$^{12+}$ impurity ions. The horizontal dashed lines are the threshold $\langle\mathcal{T}\rangle_{ thr}$, calculated as explained in the text.}
\label{FigS2}
\end{figure}

For a given impurity content (fixed $N_{ Xe}$, hence fixed $Q_{imp}$), we then determine $\Gamma_c$ as the crossing of $\langle\mathcal{T}\rangle_{ thr}$. If the crossing occurs between neighboring points $(\Gamma_i,\langle\mathcal{T}\rangle_i)$ and $(\Gamma_{i+1},\langle\mathcal{T}\rangle_{i+1})$ satisfying
\begin{equation}
\langle\mathcal{T}\rangle_i < \langle\mathcal{T}\rangle_{ thr} \le \langle\mathcal{T}\rangle_{i+1},
\end{equation}
we define $\Gamma_c$ by linear interpolation,
\begin{equation}
\Gamma_c = \Gamma_i + (\Gamma_{i+1}-\Gamma_i)\,
\frac{\langle\mathcal{T}\rangle_{ thr}-\langle\mathcal{T}\rangle_i}{\langle\mathcal{T}\rangle_{i+1}-\langle\mathcal{T}\rangle_i}.
\end{equation}
If no crossing is observed within the scanned range, the corresponding dataset is excluded from the $\Gamma_c$ extraction for that impurity setting. An example is shown in Fig. S1.

We estimate the uncertainty on $\Gamma_c$ by Monte Carlo propagation of the measured $\langle\mathcal{T}\rangle$ noise through the full thresholding analysis. For each dataset and confinement point $\Gamma_i$, we model the measured mean sharpness as
$\langle\mathcal{T}\rangle_i^{(\alpha)}=\langle\mathcal{T}\rangle_i + (\sigma_{\mathcal{T},i}/\sqrt{n_i})\,t_{\nu_i}$,
with $\nu_i=n_i-1$, where $\sigma_{\mathcal{T},i}$ and $n_i$ are the within-group standard deviation and number of contributing frames. The dataset-specific crystallinity threshold is resampled consistently from the undoped reference in the same way. For each Monte Carlo realization we recompute $\Gamma_c$ using the same crossing criterion with linear interpolation between adjacent $\Gamma_i$ points. Finally, at fixed $Q_{imp}$ we aggregate across datasets by bootstrap resampling of datasets and take the standard deviation of the resulting $\Gamma_c$ distribution as the uncertainty. The results are shown in Fig.~S2.

\subsection*{Influence length from spatial Tenengrad maps}
\label{sec:influence}

To estimate the spatial range over which an embedded Xe$^{12+}$ ion perturbs the Ca$^+$ host, we analyze the spatial distribution of the Tenengrad energy map $\mathcal{T}(x,y)$ defined above. The analysis is performed along the main (axial) direction of the crystal, denoted by $x$.

For each fluorescence frame we compute a macrocell Tenengrad map by dividing the image into non-overlapping square macrocells and averaging $\mathcal{T}(x,y)$ within each macrocell. We use macrocells to suppress pixel-scale noise while retaining spatial structure on the scale of the dopant-induced perturbation. If the macrocells are chosen too small, $\Delta\mathcal{T}(x,y)$ is dominated by noise; if they are chosen too large, the local perturbation is averaged out and the extracted range is biased low. We therefore use square macrocells of side 32 pixels as a compromise, and verified that the extracted influence length is stable over a broad range of macrocell sizes around this value.

\begin{figure}
\centering
\includegraphics[width=0.475\textwidth]{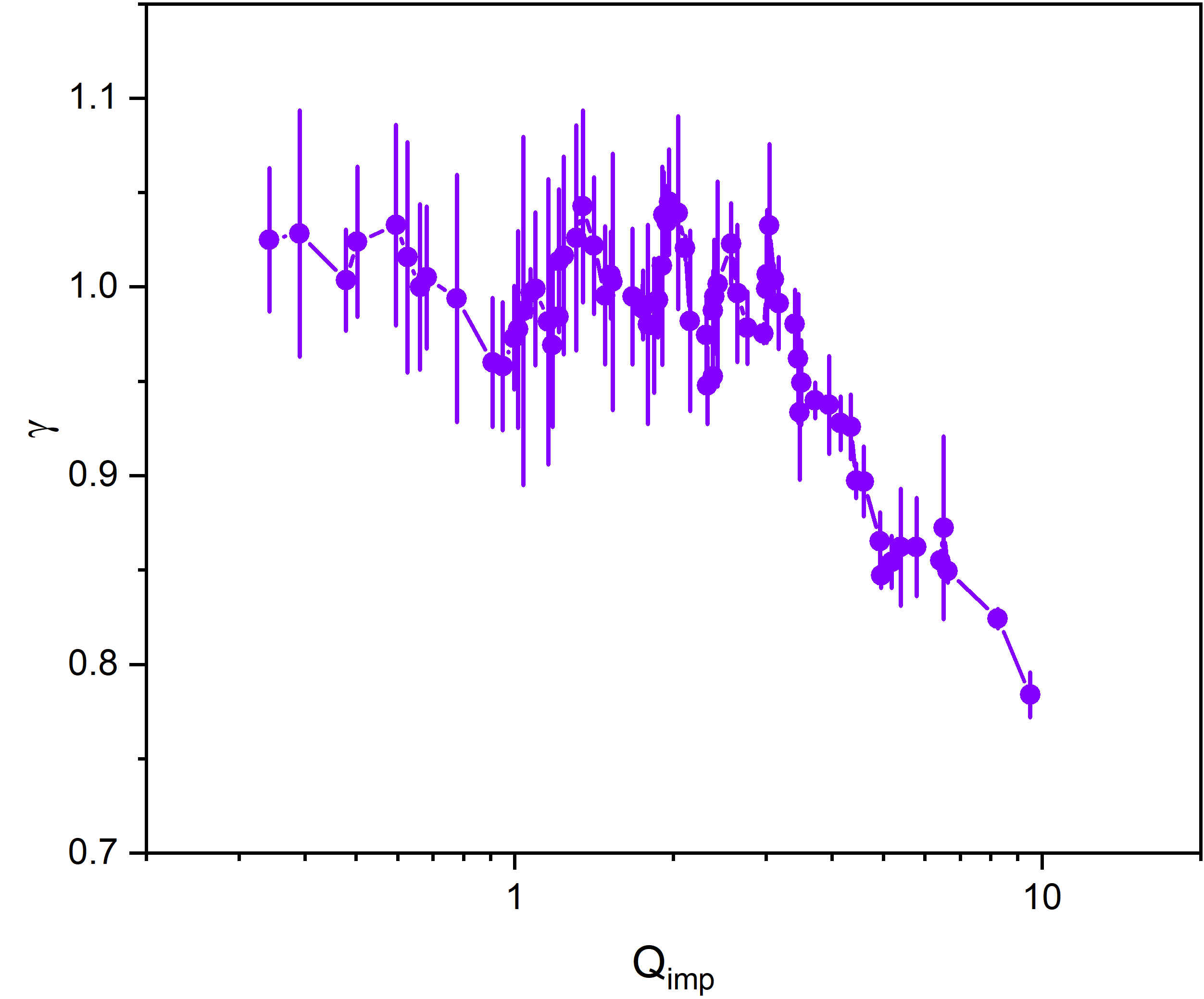}
\caption{Normalized critical Coulomb parameter $\gamma$ as
a function of the impurity parameter $Q_{imp}$. The points are the same as those shown in Fig. 2 in the main text but on a horizontal logarithmic scale. The error bars are calculated with the method explained in the text.}
\label{FigS3}
\end{figure}

For a given dataset and confinement setting, we average the resulting macrocell maps over the five frames to obtain a single map $\mathcal{T}(x,y)$ for that setting. We then form a baseline-subtracted map by subtracting the corresponding undoped reference acquired under identical conditions,
\begin{equation}
\Delta\mathcal{T}(x,y)\equiv \mathcal{T}(x,y;N_{ Xe})-\mathcal{T}(x,y;N_{ Xe}=0).
\end{equation}
To obtain a one-dimensional profile, we average $\Delta\mathcal{T}(x,y)$ along the transverse direction,
\begin{equation}
\Delta\overline{\mathcal{T}}(x)\equiv \langle \Delta\mathcal{T}(x,y)\rangle_y.
\end{equation}

We define the perturbation center as the location of the maximum of $\Delta\overline{\mathcal{T}}(x)$,
\begin{equation}
x_0 \equiv { arg\,max}_x\,\Delta\overline{\mathcal{T}}(x),
\end{equation}
and estimate a noise floor from the far-field tails of the profile. Specifically, we take the first and last 20\% of the sampled $x$-range and exclude a window around $x_0$ (20\% of the sampled range) to avoid contamination from the impurity signal. From the remaining tail values we estimate the noise scale using a robust median-absolute-deviation estimator, and define a threshold
\begin{equation}
\Delta\overline{\mathcal{T}}_{ thr}\equiv 3\,\sigma_{\Delta\overline{\mathcal{T}}}.
\end{equation}
We then determine the first positions where $\Delta\overline{\mathcal{T}}(x)$ falls below $\Delta\overline{\mathcal{T}}_{ thr}$ for two consecutive sample points, yielding $\xi$.

As one would expect, the extracted influence lengths show substantial scatter across datasets, reflecting variations in both dopant content and overall crystal size: depending on the specific sample, the perturbed region can be negligible compared to the crystal length or extend over a large fraction of it. To obtain a compact, comparable indicator, we therefore group the measurements by confinement (i.e. by the same  $\Gamma$) and compute the mean influence length. This reveals two reproducible regimes: the ensemble averaged influence length is approximately 50 $\mu$m on the more liquid side, and increases to approximately 60 $\mu$m on the crystalline side, as shown in Fig. S3. While this averaging is only indicative, it is consistent with the existence of distinct pre- and post-crystallization responses to impurities, and provides an internal cross-check of the crossover assignment obtained from the Tenengrad thresholding.

\subsection*{Lindemann parameter in small crystals}
\label{sec:lindemann_sm}

For small crystals (up to $\sim 20$ Ca$^+$ ions) individual ions are reliably resolved in the fluorescence images. This enables an ion-resolved estimate of the Lindemann parameter,
\begin{equation}
\mathcal{L}=\frac{\sqrt{\langle u^2\rangle}}{a},
\end{equation}
where $\langle u^2\rangle$ is the mean-squared displacement about equilibrium sites and $a$ is a characteristic inter-ion spacing. In practice, for a finite ion chain/crystal observed by imaging, we extract a robust per-frame estimator of $\mathcal{L}$ from the measured ion positions and spot sizes, as described below.

Each fluorescence frame is first bandpass-filtered to suppress slowly varying background and enhance ion-like features. We use a difference of Gaussians,
\begin{equation}
G(x,y)=\left[\,\mathcal{G}_{\sigma_s}*I\,\right](x,y)-\left[\,\mathcal{G}_{\sigma_l}*I\,\right](x,y),
\end{equation}
where $I(x,y)$ is the raw image, $\mathcal{G}_\sigma$ denotes a Gaussian kernel of width $\sigma$, and $*$ is discrete convolution. We then identify ion candidates as local maxima of $G(x,y)$ with a minimum separation (to avoid double-counting) and a relative height threshold referenced to the maximum of $G$ in that frame. For each candidate peak at $(x_0,y_0)$ we refine the centroid and estimate a spot width by computing second moments in a local patch. Specifically, within a square patch of half-width $w$ pixels we subtract a local background (median of the patch), clip negative values, and compute intensity-weighted centroids $(\bar{x},\bar{y})$ and variances $(\sigma_x^2,\sigma_y^2)$. We define a single spot-size parameter
\begin{equation}
\sigma \equiv \sqrt{\frac{\sigma_x^2+\sigma_y^2}{2}}.
\end{equation}
Peaks that yield unphysical $\sigma$ values (too small or too large) are discarded. The resulting list of ion centers $\{\mathbf{r}_i\}=\{(\bar{x}_i,\bar{y}_i)\}$ and spot sizes $\{\sigma_i\}$ constitutes the per-frame ion-resolved data. From the extracted ion centers we compute a nearest-neighbour spacing for each ion and then average over ions,
\begin{equation}
d_{ nn}\equiv \left\langle \min_{j\neq i}|\mathbf{r}_i-\mathbf{r}_j| \right\rangle_i.
\end{equation}
This provides an estimate of the characteristic inter-ion spacing in the image plane. For the localization length, we use the median spot size,
\begin{equation}
\sigma_{ med}\equiv { median}\{\sigma_i\}.
\end{equation}
We then define the Lindemann estimator as
\begin{equation}
\mathcal{L}= \frac{\sigma_{ med}}{d_{ nn}}.
\end{equation}
This quantity tracks the ratio of a typical positional spread (spot width) to the inter-ion spacing, and is robust against occasional missed or spurious peaks due to the use of medians and nearest-neighbour distances. For each confinement setting we evaluate $\mathcal{L}$ over the available frames and compute the mean and standard deviation.

\begin{figure}
\centering
\includegraphics[width=0.475\textwidth]{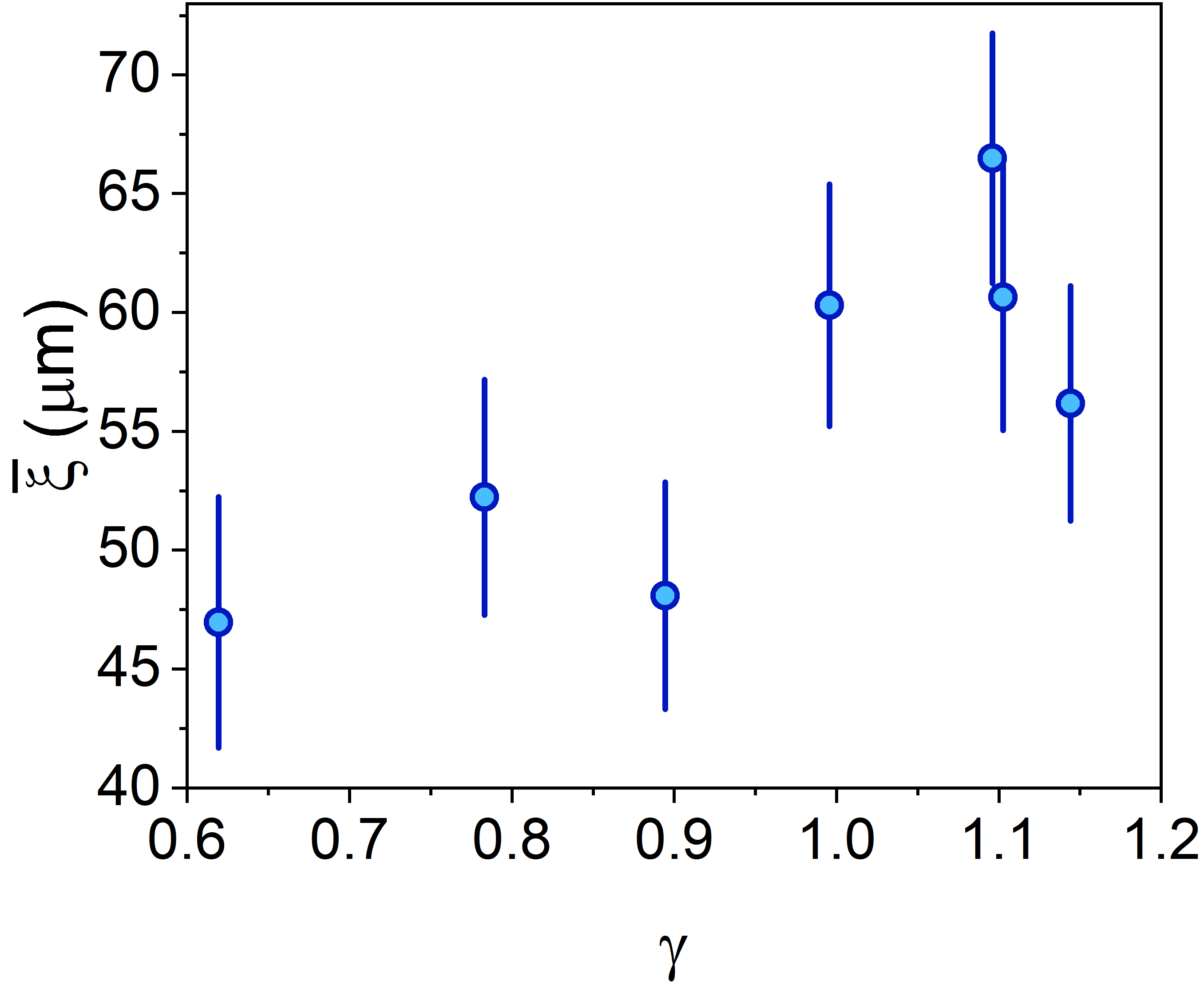}
\caption{Mean influence length $\bar{\xi}$ extracted from axial profiles as a function of the normalized crystallization threshold $\gamma$. Each point shows the average over all samples measured at a given confinement setting (fixed $\Gamma$), and error bars indicate the corresponding standard deviation across the ensemble.}
\label{FigS1}
\end{figure}

The measured spot size $\sigma_i$ includes contributions from the imaging point-spread function, residual micromotion blur, and any out-of-plane excursions projected into the image. These contributions primarily add an approximately confinement-independent floor to $\sigma_{ med}$ for a fixed imaging configuration. Since the analysis focuses on how $\mathcal{L}$ changes with confinement and impurity content under identical optical conditions, and since $d_{ nn}$ is extracted from the same frames, the resulting systematic largely cancels in relative comparisons. 

\subsection*{Scalings in stellar models}
\label{sec:astro_sm}

The relations below are derived in the classical Coulomb (OCP) limit (weak screening) and are intended as transparent scalings rather than a self-consistent stellar-evolution calculation. In particular, we do not include electron-screening corrections to the interaction, multicomponent phase-diagram effects (phase separation, distillation, sedimentation), or additional energy sources. As discussed in the main text, for finite screening one replaces the OCP freezing condition with the Yukawa melting curve $\Gamma_m(\kappa)$ and then applies the same multiplicative correction factor $\gamma(Q_{imp})$ \cite{Hamaguchi1997,Khrapak2015}.

At fixed density (fixed $a$) one has $\Gamma\propto 1/T$. If crystallization is triggered by $\Gamma=\Gamma_c$, then
\begin{equation}
\frac{T_m(Q_{imp})}{T_m(0)}=\frac{1}{\gamma(Q_{imp})},
\end{equation}
where $T_m$ is the local melting (crystallization) temperature at the relevant depth. In white dwarfs, for a degenerate core with an insulating envelope, Mestel cooling gives the scaling of the stellar luminosity $\mathcal{L}$ with core temperature $T_c$,
\begin{equation}
\mathcal{L}\propto T_c^{7/2},
\end{equation}
and the cooling age $t$ with luminosity,
\begin{equation}
t\propto \mathcal{L}^{-5/7},
\end{equation}
see \cite{Mestel1952,ShapiroTeukolsky}. Identifying the onset of crystallization at the depth where $T_c\simeq T_m$, the impurity-induced shift implies
\begin{equation}
\frac{\mathcal{L}(Q_{imp})}{\mathcal{L}(0)}
=\left(\frac{T_m(Q_{imp})}{T_m(0)}\right)^{7/2}
=\left(\frac{1}{\gamma(Q_{imp})}\right)^{7/2},
\end{equation}
and the corresponding shift of the crystallization age $t_{ crys}$ (the age at which that depth reaches $T_c\simeq T_m$) scales as
\begin{equation}
\frac{t_{ crys}(Q_{imp})}{t_{ crys}(0)}
=\left(\frac{\mathcal{L}(Q_{imp})}{\mathcal{L}(0)}\right)^{-5/7}
=\gamma(Q_{imp})^{5/2}.
\end{equation}
These relations are the ones used for Fig.~4(a,b) of the main text. To realize Fig. 4, we parameterize the crossover observed in Fig.~2 of the main text with a minimal knee model, consisting of a low-$Q_{imp}$ plateau followed by a linear decrease,
\begin{equation}
\gamma(Q_{imp}) = 1 + s\,\max\!\left(0,\,Q_{imp}-c\right),
\end{equation}
where $c$ is the knee position and $s$ is the slope in the high-$Q_{imp}$ regime. A weighted least-squares fit yields $s = -0.033 \pm 0.002$ and $c = 2.1 \pm 0.2$.

\end{document}